\pdfoutput=1

\documentclass[11pt,twoside,a4paper,cmspaper,final,collab]{cms-tdr}
\def\svnVersion{442240P}\def\cmsCernNoTag{CERN-EP-2016-309}\def\cmsCernDate{\today}\def\cmsMessage{Published in the Journal of High Energy Physics as \href{http://dx.doi.org/10.1007/JHEP03(2017)061}{\doi{10.1007/JHEP03(2017)061.}}}

\begin{document}\cmsNoteHeader{EXO-16-010}

\hyphenation{had-ron-i-za-tion}
\hyphenation{cal-or-i-me-ter}
\hyphenation{de-vices}
\RCS$Revision: 435446 $
\RCS$HeadURL: svn+ssh://svn.cern.ch/reps/tdr2/papers/EXO-16-010/trunk/EXO-16-010.tex $
\RCS$Id: EXO-16-010.tex 435446 2017-11-21 08:49:51Z aalbert $

\newcommand{\ZZ}{\ensuremath{\cPZ\cPZ}\xspace}
\newcommand{\WZ}{\ensuremath{\PW\cPZ}\xspace}
\newcommand{\WW}{\ensuremath{\PW\PW}\xspace}
\newcommand{\Wjets}{\ensuremath{\PW\text{+jets}}\xspace}
\renewcommand{\Z}{\ensuremath{\cPZ}\xspace}
\newcommand{\W}{\ensuremath{\PW}\xspace}
\newcommand{\tw}{\ensuremath{\cPqt\PW}\xspace}
\newcommand{\dyll}{\ensuremath{\Z/\gamma^*\to\ell^+\ell^-}\xspace}
\providecommand{\cmsLeft}{left}
\providecommand{\cmsRight}{right}
\providecommand{\CL}{CL\xspace}
\providecommand{\NA}{\ensuremath{\text{---}}\xspace}

\cmsNoteHeader{EXO-16-010}
\title{Search for dark matter and unparticles in events with a \Z boson and missing transverse momentum in proton-proton collisions at $\sqrt{s} = 13 \TeV$}

\date{\today}

\abstract{A search for dark matter and unparticle production at the LHC has been performed using events
containing two charged leptons (electrons or muons), consistent with the decay of a Z boson, and large missing transverse momentum.
This study is based on data collected with the CMS detector in 2015, corresponding to
an integrated luminosity of 2.3\fbinv of proton-proton collisions at the LHC, at a center-of-mass energy of 13\TeV.
No excess over the standard model expectation is observed.
Compared to previous searches in this topology, which exclusively relied on effective field theories,
the results are interpreted in terms of a simplified model of dark matter production for both vector and axial vector couplings between a mediator and dark matter particles.
The first study of this class of models using CMS data at $\sqrt{s}=13$\TeV is presented.
Additionally, effective field theories of dark matter and unparticle production are used to interpret the data.
}

\hypersetup{%
pdfauthor={CMS Collaboration},%
pdftitle={Search for dark matter and unparticles in events with a Z boson and missing transverse momentum in proton-proton collisions at sqrt(s) = 13 TeV},%
pdfsubject={CMS},%
pdfkeywords={CMS, physics, exotica, dark matter, unparticles}}

\maketitle

\section{Introduction}
\label{sec:intro}

According to the well-established $\Lambda_\mathrm{CDM}$ model of cosmology, known matter only comprises about 5\% of the total energy content of the universe, with 27\% contributed by dark matter (DM) and the rest being dark energy~\cite{Ade:2015xua}.
Although strong astrophysical evidence indicates the existence of DM,
there is no evidence yet for nongravitational interactions between DM and standard model (SM) particles.
DM searches exploit a number of methods including direct detection~\cite{Cushman:2013zza} and indirect detection~\cite{Buckley:2013bha}.
If there are DM particles that can be observed in direct detection experiments,
they could have substantial couplings to nucleons, and therefore could be produced at the CERN LHC.
A theoretically promising possibility is that DM may take the form of weakly interacting massive particles. Searches for production of such particles at colliders typically consider the case of DM recoiling against a standard model particle (``tag'') to obtain a defined signature~\cite{Abercrombie:2015wmb}. Such searches have been performed using various standard model signatures as tags~\cite{Aad:2013oja,Aad:2014vka,ATLAS:2014wra,Khachatryan:2014tva,Aad:2014vea,Aad:2014tda,Khachatryan:2014uma,Aad:2014wza,Aad:2015yga,Khachatryan:2015nua,Khachatryan:2014rra,Aad:2015zva,Khachatryan:2015bbl,Khachatryan:2014rwa,Khachatryan:2016mdm,Aaboud:2016qgg}. In models where DM production is mediated by an interaction involving SM quarks, the monojet signature is typically the most sensitive. If DM particles are instead produced via radiation emitted by a standard model boson, searches in the $\Z/\W/\gamma+\ETm$ channels are advantageous.

The study presented here considers the case of a \Z boson recoiling against a pair of DM particles, $\chi\overline\chi$. The \Z boson subsequently decays into
two charged leptons ($\ell^{+}\ell^{-}$, where $\ell=\Pe$ or $\Pgm$) producing a well-defined signature together with missing transverse momentum
due to the undetected DM particles.
A simplified tree-level ultraviolet-complete model~\cite{Abercrombie:2015wmb} that contains
a massive spin-1 mediator exchanged in the $s$-channel is considered here.
In this model, the spin-1 mediator ${\cal A}$ could have either vector or axial-vector couplings to the SM and DM particles. The DM particle $\chi$ is assumed to be a Dirac fermion.
The interaction Lagrangian of the $s$-channel vector mediated DM model can be written as:
\begin{equation*}
\mathcal{L}_{\text{vector}} = - \sum_{q} g_{\Pq} \mathcal{A}_{\mu} \overline{q}\gamma^{\mu}q - g_{\chi} \mathcal{A}_{\mu}\overline{\chi}\gamma^{\mu}\chi,
\end{equation*}
where the mediator is labeled as $\mathcal{A}$, and its coupling to DM particles is labeled as $g_\chi$.
The coupling between the mediator and SM quarks is labeled as $g_{\Pq}$, and is assumed to be universal to all quarks.
The Lagrangian for an axial-vector mediator is obtained by making the replacement $\gamma^{\mu}\to\gamma^{\mu} \gamma^{5}$ in all terms.

As a benchmark model for DM production via a scalar coupling, an effective field theory (EFT) with dimension-7 operators is also considered~\cite{Abercrombie:2015wmb}. It contains ${SU}(2)_\mathrm{L}  \times {U}(1)_\mathrm{Y}$ gauge invariant couplings between a DM pair and two SM gauge bosons in a four-particle contact interaction. The corresponding interaction Lagrangian is:
\begin{equation*}
\mathcal{L}_\text{dim. 7} =  \frac{1}{\Lambda^3} \overline\chi \chi \left( c_{1} B_{\mu\nu} B^{\mu\nu} + F^{i}_{\mu\nu} F^{i, \mu\nu} \right),
\label{eq:Lag_DMVV}
\end{equation*}
in which $B_{\mu\nu}$ and $F^{i}_{\mu\nu}$ are the $U(1)_{Y}$ and $SU(2)_{L}$ field tensors, and $\Lambda$ denotes the cutoff scale.
The coupling parameter $c_{1}$ controls the relative importance of the $U(1)_{Y}$ and $SU(2)_{L}$ fields for DM production.
Any multiplicative factor for the ${U}(1)_{Y}$ and ${SU}(2)_{L}$ couplings is absorbed into $\Lambda$. Note that the choice of $\Lambda$ modifies the signal cross section, but not the expected kinematic properties of events. The model is nonrenormalizable and should be considered as a benchmark of the sensitivity to this class of interaction. It should be used with caution when making comparisons with other sources of DM constraints, such as direct detection experiments.

\begin{figure}[h!tb]
\centering
\includegraphics[width=0.45\textwidth]{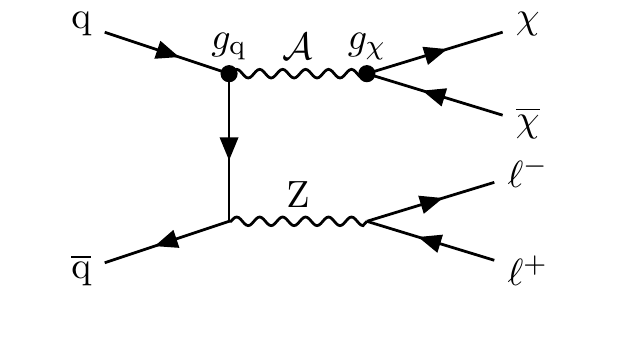}
\includegraphics[width=0.45\textwidth]{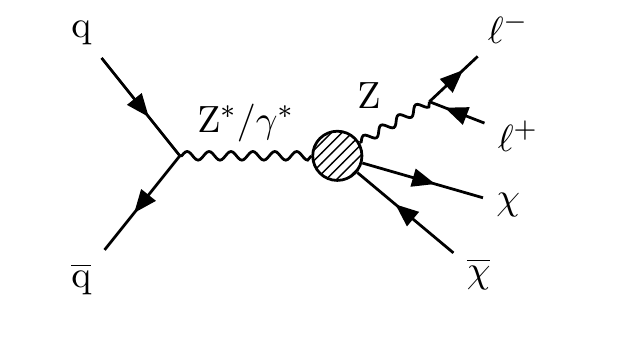}
\caption{Leading order Feynman diagrams for production of DM pairs  ($\chi\overline{\chi}$) in association with a \Z boson.
Left: the simplified model containing a spin-1 mediator $\mathcal{A}$. The constant $g_\Pq$ ($g_\chi$) is the coupling strength between $\mathcal{A}$ and quarks (DM). Right: an EFT benchmark with a DM pair coupling to gauge bosons via dimension-7 operators.}
\label{fig:FeynDiagrams}
\end{figure}

Figure~\ref{fig:FeynDiagrams} shows the Feynman diagrams for production of DM pairs  ($\chi\overline{\chi}$)
in association with a \Z boson in these two types of models.

The signature for DM production considered in this paper  is the production of a pair of leptons ($\Pep\Pem$ or $\Pgmp\Pgmm$)
consistent with a \Z boson decay, together with a large missing transverse momentum.
This same signature is sensitive to other models of physics beyond the SM (BSM), e.g. ``unparticles''($\textsf{U}$).

The unparticle physics concept~\cite{Georgi:2007ek,Georgi:2007si,Cheung:2007zza,Cheung:2007ap} is particularly
interesting because it is based on scale invariance, which is anticipated in many BSM physics scenarios~\cite{Kang:2014cia,Rinaldi:2014gha,Cheng:1988zx}.
The effects of the scale invariant sector (unparticles) appear as a noninteger number of invisible massless particles.
In this scenario, the SM is extended by introducing a scale invariant Banks--Zaks (BZ) field, which has a nontrivial infrared fixed point~\cite{Banks:1981nn}.
This field can interact with SM particles by exchanging heavy particles with a
high mass scale $M_{\textsf{U}}$. Below this mass scale, the coupling is nonrenormalizable and the interaction
is suppressed by powers of $M_{\textsf{U}}$.
The EFT Lagrangian can be expressed as:
\begin{equation*}
\mathcal L_{\textsf{U}} = C_{\textsf{U}}
                \frac{\Lambda_{\textsf{U}}^{d_\mathrm{BZ}-d_{\textsf{U}}}}
                {M_{\textsf{U}}^{k}}\mathcal{O}_{\mathrm{SM}}\mathcal{O}_{\textsf{U}}
        = \frac{\lambda}{\Lambda_{\textsf{U}}^{d_{\textsf{U}}}} \mathcal{O}_{\mathrm{SM}}\mathcal{O}_{\textsf{U}},
\label{eq:Lagunparticle}
\end{equation*}
in which $C_{\textsf{U}}$ is a normalization factor, $d_{\textsf{U}}$
represents the possible noninteger scaling dimension of the unparticle operator $\mathcal{O}_{\textsf{U}}$, $\mathcal{O}_\mathrm{SM}$ is an operator composed of SM fields with dimension $d_\mathrm{SM}$,
 $k=d_{\mathrm{SM}}+d_\mathrm{BZ}-4>0$ is the scaling dimension, $\Lambda_{\textsf{U}}$ is the energy scale of the interaction, and $d_\mathrm{BZ}$ denotes the scaling dimension of the BZ operator at energy scales above $\Lambda_{\textsf U}$. The parameter $\lambda = C_{\textsf{U}}\Lambda^{d_\mathrm{BZ}}_{\textsf{U}}/M_{\textsf{U}}^{k}$
is a measure of the coupling between SM particles and unparticles. The scaling dimension $d_{\textsf{U}}\geq1$ is constrained by the unitarity condition.
Additional details regarding this unparticle model are available in Ref.~\cite{Khachatryan:2015bbl}.

\begin{figure}[h!tb]
\centering
   \includegraphics[width=0.35\textwidth]{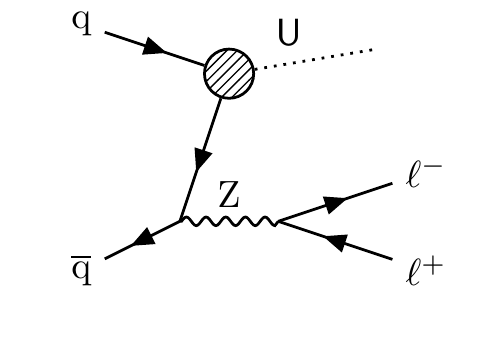}
  \caption{Leading order Feynman diagram for unparticle (denoted by {\sffamily U}) production in association with a \Z  boson.
           The hatched circle indicates the interaction modeled with an EFT operator.}
\label{fig:Unpart}
\end{figure}

In this paper, real emission of scalar unparticles is considered. The 
unparticles are assumed to couple to the standard model quarks in an 
effective three-particle interaction. In the scalar unparticle case, $\mathcal{O}_\mathrm{SM} = \overline{q}q$, which 
yields numerically identical results to the pseudo-scalar operator 
choice  $\mathcal{O}_\mathrm{SM} = \overline{q}i\gamma_{5}q$~\cite{Ask:2009pv}. Figure~\ref{fig:Unpart} shows the corresponding 
tree-level diagram for the production of unparticles associated with a Z 
boson.

The analysis is based on a data set recorded with the CMS detector in 2015 in pp collisions at a center-of-mass energy of 13\TeV, corresponding to an integrated luminosity of $2.3\pm 0.1$\fbinv. A previous CMS search in the same final state~\cite{Khachatryan:2015bbl},
based on data collected at a center-of-mass energy of 8\TeV, found no evidence of new physics and set limits on DM and unparticle
production using an EFT description. A CMS analysis of the 8\TeV data set in the combined monojet and hadronic mono-V (where V = \W or \Z) channels~\cite{Khachatryan:2016mdm} has previously set limits on the simplified model parameters considered here. Dark matter particle masses of up to 500 \GeV (400\GeV) and mediator masses of up to 1.6\TeV have been excluded in the vector (axial-vector) coupling scenarios for $g_{\Pq}=g_\mathrm{DM}=1$.
A search performed by the ATLAS Collaboration using $\sqrt{s}=13\TeV$ data corresponding to an integrated luminosity of 3.2\fbinv in events with a hadronically decaying V boson and \ETm has recently reported exclusion of the dimension-7 EFT scenario up to $\Lambda=700\GeV$ (460\GeV) for DM particle masses of 1\GeV (1\TeV)~\cite{Aaboud:2016qgg}.

\section{The CMS detector}
\label{sec:cms}

The central feature of the CMS apparatus is a superconducting solenoid of 6\unit{m} internal diameter, providing
a magnetic field of 3.8\unit{T}. Within the solenoid volume are a silicon pixel and strip tracker, a
lead tungstate crystal electromagnetic calorimeter (ECAL), and a brass and scintillator hadron calorimeter (HCAL),
each composed of a barrel and two endcap sections. Forward calorimeters extend the pseudorapidity ($\eta$)~\cite{CMSdetector}
coverage provided by the barrel and endcap detectors. Muons are measured in gas-ionization detectors embedded in the
steel flux-return yoke outside the solenoid.
A more detailed description of the CMS detector, together with a definition of the coordinate
system used and the relevant kinematic variables, can be found in Ref.~\cite{CMSdetector}.

Variables of particular relevance to the present analysis are the missing transverse momentum
vector \ptvecmiss and the magnitude of this quantity, \ETm. The quantity \ptvecmiss is defined as
the projection on the plane perpendicular to the beams of the negative vector sum of the momenta of
all reconstructed particles in an event.

\section{Simulation}
\label{sec:simulation}
\sloppy
Samples of simulated DM particle events for both the simplified model and EFT interpretations
 are generated using \MADGRAPH{}5\_a\MCATNLO 2.2.2~\cite{Alwall:2014hca} at leading order (LO) and matched to
\PYTHIA8.205~\cite{Sjostrand:2007gs} using tune CUETP8M1 for parton showering and hadronization~\cite{Khachatryan:2015pea,Skands:2014pea}.
The factorization and re\-nor\-ma\-li\-zat\-ion scales are set to the geometric mean of $\sqrt{\smash[b]{\pt^2+m^2}}$ for all final-state particles~\cite{Alwall:2014hca,Abercrombie:2015wmb}, where $\pt$ and $m$ are the transverse momentum and mass of each particle.

For the simplified model of DM production, couplings are chosen according to the recommendations in Ref.~\cite{Boveia:2016mrp}.
The coupling $g_{\chi}$ is set to unity. For $g_{\Pq}$, values of $1.0$ and $0.25$ are considered. The width of the mediator is assumed to be determined exclusively by the contributions from the couplings to quarks and the DM particle $\chi$. Under this assumption, the width is in the range 1--5\% (30--50\%) of the mediator mass for $g_{\Pq}=0.25$ ($g_{\Pq}=1.00$).
The signal simulation samples with $g_{\Pq}=1.0$ are processed using the detector simulation described below.
Signal predictions for $g_{\Pq}=0.25$ are obtained by applying event weights based on the \ETm distribution at the generator level to the fully simulated samples with $g_{\Pq}=1.0$.
This procedure allows to take into account any effect of the coupling dependent mediator width on the \ETm distribution~\cite{Boveia:2016mrp}.  The exact dependence of the width on the model parameters is reported in~\cite{Boveia:2016mrp}.

Samples for the EFT DM benchmark are generated with $\Lambda=3\TeV$ and $c_{1}=1$. Signal predictions for other values of $\Lambda$ are obtained by rescaling the signal cross section accordingly, while other values of $c_{1}$ are evaluated using the same reweighting method as for the simplified model case.

The events for the unparticle
model are generated at LO with \PYTHIA8~\cite{Ask:2008fh,Ask:2009pv}
assuming a cutoff scale $\Lambda_{\textsf{U}}=15\TeV$,
using tune CUETP8M1 for parton showering and
hadronization.
We evaluate other values of $\Lambda_{\textsf{U}}$ by rescaling the cross sections as needed.
The parameter $\Lambda_{\textsf{U}}$ acts solely as a scaling factor
for the cross section and does not influence the kinematic distributions of unparticle production~\cite{Ask:2009pv}.

\begin{figure}[htb]
\centering
\includegraphics[width=0.48\textwidth]{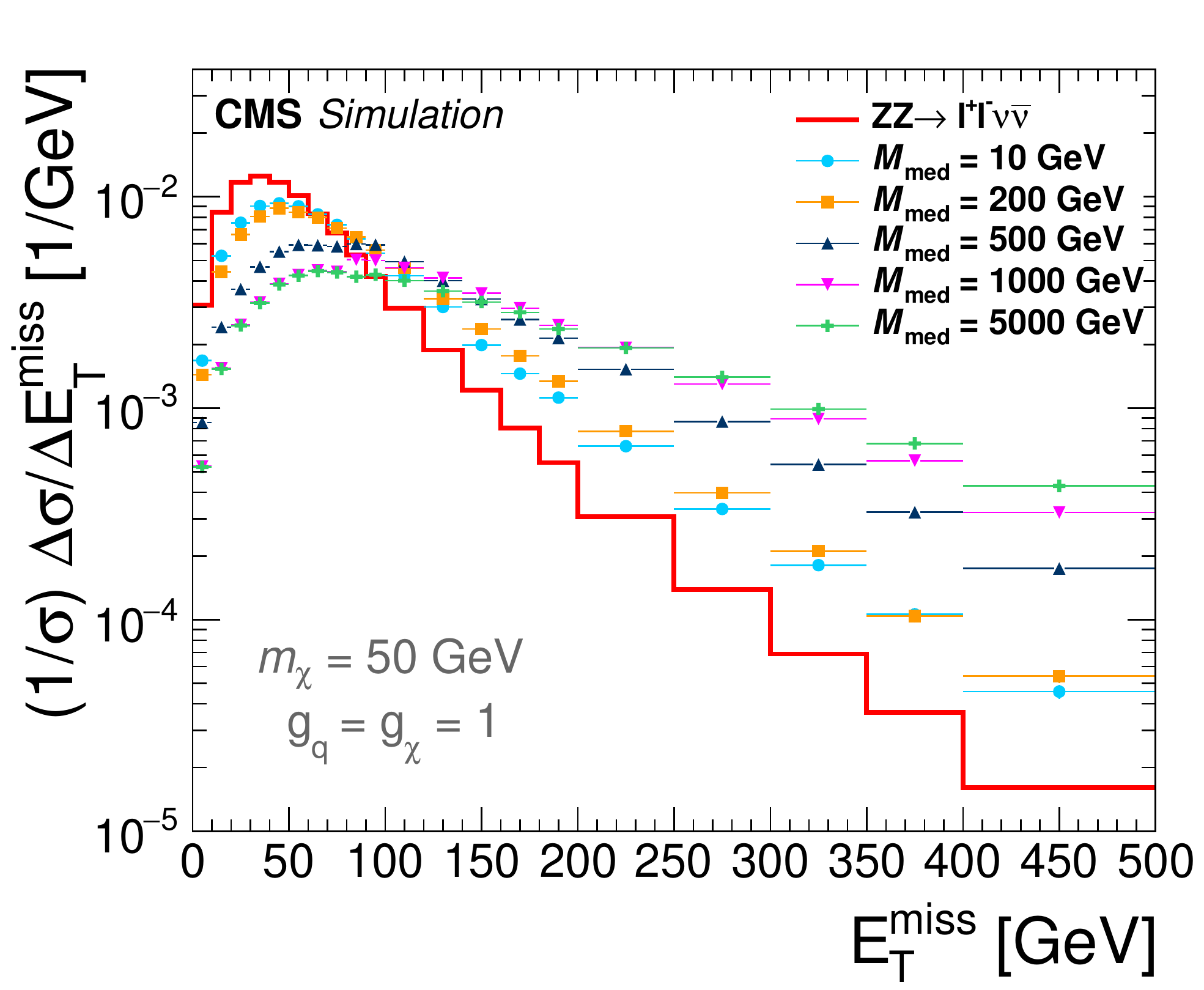}
\includegraphics[width=0.48\textwidth]{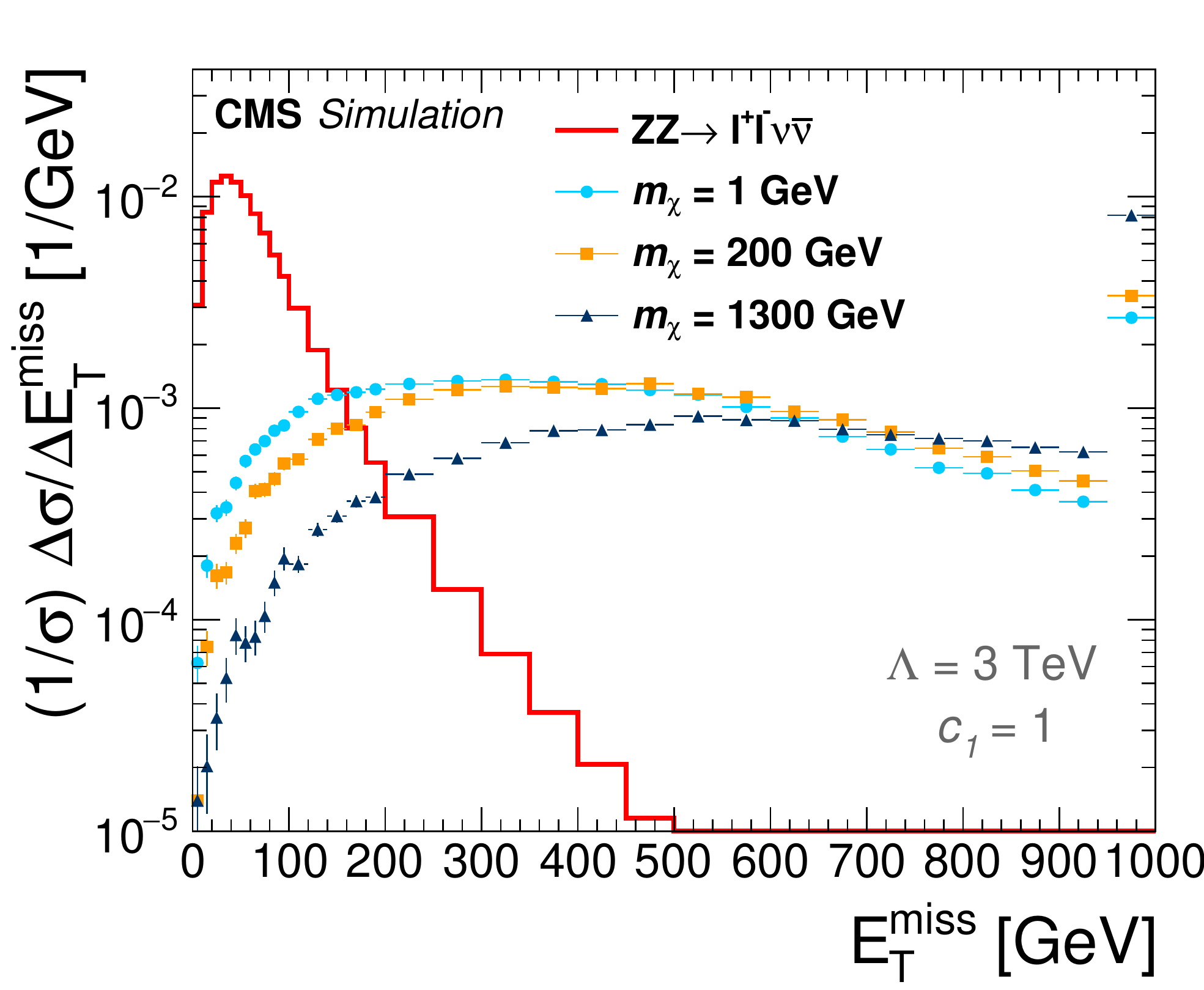}
\includegraphics[width=0.48\textwidth]{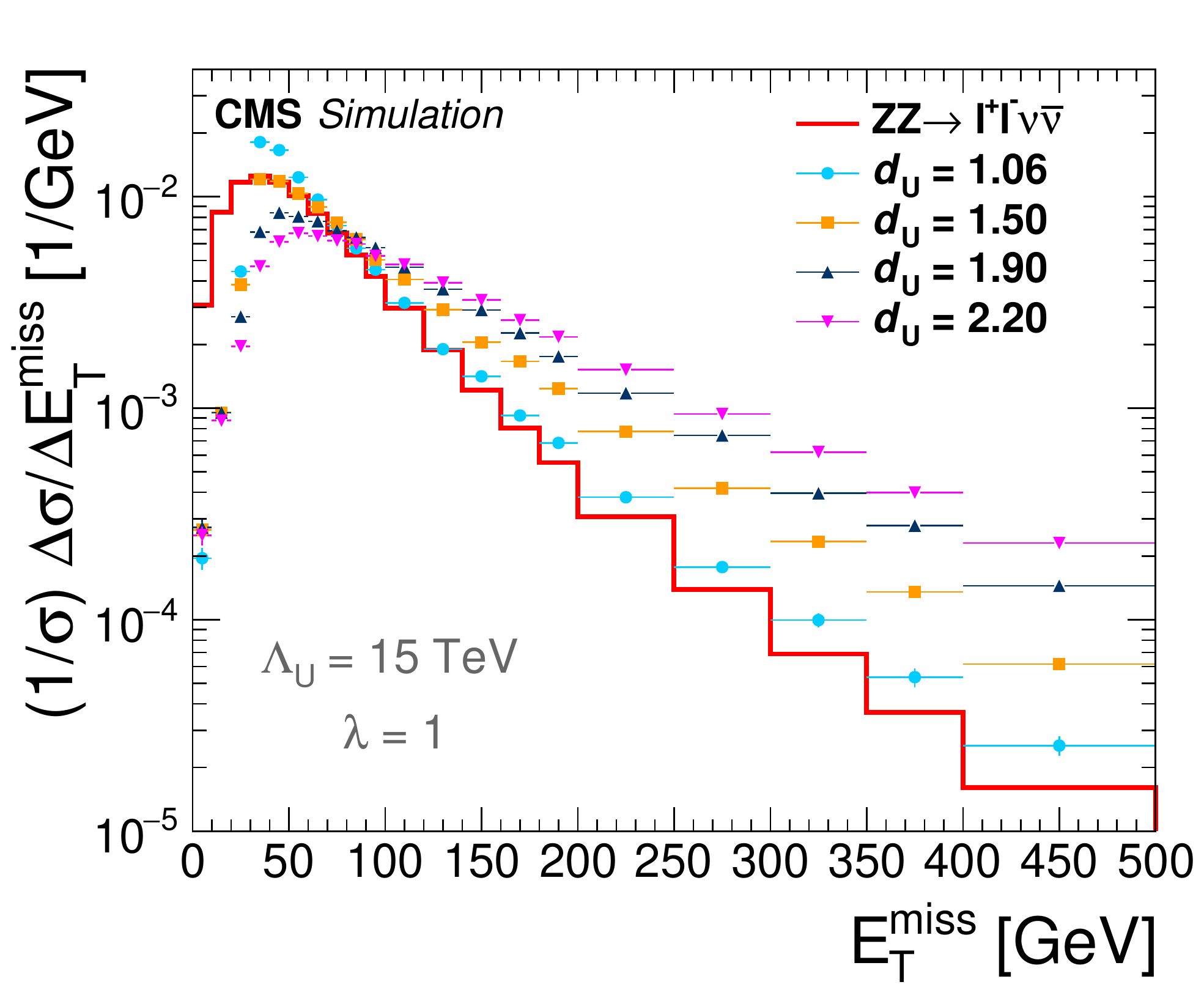}
\caption{The distribution in \ETm at the generator
        level, for the simplified DM model with vector mediator (upper \cmsLeft), EFT DM model (upper \cmsRight), and unparticle scenarios (lower panel).
        The y-axis corresponds to the integrated cross section per bin divided by the total cross section and bin width.
        The DM curves are shown for different values of the vector mediator mass $M_\text{med}$ in the upper left panel and for different values of the DM mass
        $m_\chi$ in the upper right panel.
        The unparticle curves have the scalar unparticle coupling $\lambda$ between unparticle and SM
        fields set to 1. They are shown for several values of the scaling dimension $d_{\textsf{U}}$ ranging from 1.06 to 2.20,
		spanning the region of sensitivity of this analysis.
        The SM background $\Z\Z\to\ell^{-}\ell^{+}\PGn\PAGn$ is shown as a red solid histogram. The rightmost bins include overflow.
}
\label{fig:genlevel_met}
\end{figure}

The \POWHEG2.0~\cite{Nason:2004rx,Frixione:2007vw,Alioli:2010xd,Re:2010bp,Alioli:2011as} event generator is used to produce samples of
events for the $\ttbar$, $\tw$, $\PQq\PAQq\to\ZZ$, and $\WZ$ background processes, which are simulated at next-to-leading order (NLO).
The $\Pg\Pg\to\ZZ$ process is simulated using \MCFM 7.0.1~\cite{MCFM} at NLO.
The Drell--Yan (DY, $\dyll$) process is generated using the \MADGRAPH{}5\_a\MCATNLO event generator at LO
and normalized to the next-to-next-to-leading order (NNLO) cross section as calculated using {\FEWZ} 3.1~\cite{Gavin:2010az,Li:2012wna}.
Triboson events (WZZ, WWZ and ZZZ) are simulated using \MADGRAPH{}5\_a\MCATNLO at NLO.
Samples of quantum chromodynamics (QCD) production of multijet events are generated using \PYTHIA8 at LO.
For all SM simulation samples, parton showering and hadronization are performed with \PYTHIA8 with tune CUETP8M1.

The parton distribution function (PDF) set NNPDF3.0~\cite{Ball:2014uwa} is used for Monte Carlo (MC) samples, and the detector response is simulated using a detailed description of the CMS detector, based on the \GEANTfour
package~\cite{Agostinelli:2002hh,Allison:2006ve}. Minimum bias events are superimposed on the
simulated events to emulate the effect of additional pp interactions in the same or nearby bunch crossings (pileup).
All MC samples are corrected to reproduce the
pileup distribution as measured in the data. The average number of pileup interactions
per proton bunch crossing is about 12 for the 2015 data sample.

The upper left panel of Fig.~\ref{fig:genlevel_met} shows
the distribution of \ETm at the generator level for DM particles with a mass of 50 \GeV in the simplified model.
The events generated with larger mediator mass $M_\text{med}$ tend to have a broader \ETm distribution and reach further into the high-\ETm regime.
The analogous distributions in the EFT benchmark model with DM masses $m_\chi = 1$, 200, and 1300\GeV are shown in the upper right panel of Fig.~\ref{fig:genlevel_met}.
In the unparticle scenario, the events generated with larger scaling dimension $d_{\textsf{U}}$ tend to preferentially populate the high-\ETm regime,
as shown in the lower panel of Fig.~\ref{fig:genlevel_met}.
The SM background $\Z\Z\to\ell^{-}\ell^{+}\PGn\PAGn$ is shown in all plots for comparison, as a red solid histogram.

\section{Event reconstruction}
\label{sec:object}

Events are collected by requiring dilepton triggers ($\Pe\Pe$ or $\Pgm\Pgm$) with a threshold of $\pt>17\GeV$ for the leading lepton.
The threshold for the subleading lepton is $\pt>12\,(8)\GeV$ for electrons (muons).
Single-lepton triggers with thresholds of $\pt > 23\,(20)\GeV$ for electrons (muons)
are also included to recover residual trigger
inefficiencies. Prior to the selection of leptons, the primary vertex~\cite{Khachatryan:2010pw} with
the largest value of $\sum \pt^2$ for the associated tracks is selected as the event vertex.
Simulation studies show that this requirement
correctly selects the event vertex in more than 99\% of
both signal and background events. The lepton candidate tracks are
required to be compatible with the event vertex.

A particle-flow (PF) event algorithm~\cite{CMS-PAS-PFT-09-001,CMS-PAS-PFT-10-001}
reconstructs and identifies each individual particle with an optimized combination of
information from the various elements of the CMS detector. Photon energies are
directly obtained from the ECAL measurement, corrected for zero-suppression effects~\cite{CMSdetector}.
Electron energies are determined from a combination of the electron momentum at
the event vertex as determined by the tracker, the energy of the corresponding ECAL
cluster, and the energy sum of all bremsstrahlung photons spatially compatible with
originating from the electron track. Muon momenta are obtained from the curvature
of the corresponding track. Charged hadron energies are determined from a combination
of their momentum measured in the tracker and the matching ECAL and HCAL energy deposits,
corrected for zero-suppression effects and for the response function of the calorimeters
to hadronic showers~\cite{CMS-PAS-PFT-10-001}. Finally, neutral hadron energies are obtained from the
corresponding corrected ECAL and HCAL energies.

Electron candidates are reconstructed using an algorithm that combines
information from the ECAL and the tracker~\cite{Khachatryan:2015hwa}.
To reduce the electron misidentification rate, the candidates have to satisfy
additional identification criteria that are based on the shape of the
electromagnetic shower in the ECAL.
In addition, the electron track is required to originate from the event vertex and to match the shower cluster in the ECAL.
Electron candidates with an ECAL cluster in the
transition region between ECAL barrel and endcap ($1.44 < \abs{\eta} < 1.57$) are
rejected because the reconstruction of an electron candidate in this region is not optimal.
Candidates that are identified as coming from photon conversions~\cite{Khachatryan:2015hwa} in the
detector material are explicitly removed.

Muon candidate reconstruction is based on two algorithms: in the first,
tracks in the silicon tracker are matched with at least one muon segment in any detector plane of the muon system,
and in the second algorithm, a combined fit is performed to hits in
both the silicon tracker and the muon system~\cite{Chatrchyan:2012xi}.
The muon candidates in this analysis are required to be reconstructed
with at least one of the two algorithms and to be further identified as muons by the PF algorithm.
To reduce the muon misidentification rate, additional identification
criteria are applied based on the number of spatial points measured in the tracker and in
the muon system, the fit quality of the muon track, and its consistency with the event vertex location.

Leptons produced in the decay of \Z bosons are expected to be isolated from hadronic activity in
the event. Therefore, an isolation requirement is applied based on the sum of the momenta of the PF candidates
found in a cone of radius $\Delta R= 0.4$ around each lepton.
The isolation sum is required to be smaller than 15\% (20\%) of the $\pt$ of the electron (muon).
For each electron, the mean energy deposit in the isolation cone of the electron,
coming from other pp collisions in the same bunch crossing, is estimated
following the method described in Ref.~\cite{Khachatryan:2015hwa}, and subtracted from the isolation sum.
For muon candidates, only charged tracks associated with the event vertex are included.
The sum of the \pt for charged particles not associated with the event vertex
in the cone of interest is rescaled by a factor of $0.5$, corresponding to the average neutral
to charged energy density ratio in jets, and subtracted from the isolation sum.

For the purpose of rejecting events containing $\tau$ leptons, hadronically decaying $\tau$ leptons (\tauh) are identified using the ``hadron-plus-strips"
algorithm. The algorithm identifies a jet as a \tauh candidate if a subset of the particles assigned to the jet
is consistent with the decay products of a \tauh~\cite{Khachatryan:2015dfa}.
In addition, \tauh candidates are required to be isolated from other activity in the event.

Jets are reconstructed from PF candidates
by using the anti-\kt clustering algorithm~\cite{Cacciari:2008gp}
with a distance parameter of 0.4, as implemented in the {\FASTJET}
package~\cite{Cacciari:2011ma,Cacciari:2006gp}.
Jets are identified over the full calorimeter acceptance, $\abs{\eta} < 5$.
The jet momentum is defined as
the vector sum of all particle momenta assigned to the jet, and is found in
simulation to be within 5 to 10\% of the true hadron-level momentum over the whole \pt
range and detector acceptance. An overall energy subtraction is applied to
correct for the extra energy clustered in jets due to pileup, following the procedure described
in Ref.~\cite{Khachatryan:2016kdb}.
Additional corrections to the jet energy scale and resolution are derived from
simulation, and are complemented by measurements of the energy balance in dijet and $\gamma$+jets events~\cite{Khachatryan:2016kdb}.

\section{Event selection}
\label{sec:selection}

A preselection with a large yield is used to validate the background model and is followed by a final selection
that is designed to give maximal sensitivity to the signal, as quantified by the expected limits achieved.
Preselected events are required to have exactly two well-identified, isolated leptons with the same flavor
and opposite charge ($\Pep\Pem$ or $\Pgmp\Pgmm$), each with $\pt > 20$\GeV.
The invariant mass of the lepton pair is required to be within
$\pm$10\GeV of the nominal mass of the $\Z$ boson~\cite{Agashe:2014kda}. Only electrons (muons) within the range of
$\abs{\eta}<2.5\, (2.4)$ are considered.
To reduce the background from the $\W\Z$ process where the \W boson decays leptonically, events are removed if an additional
electron or muon is reconstructed with $\pt > 10\GeV$. The event is also removed from the final selection
if a \tauh candidate is reconstructed with $\pt>20\GeV$. As a loose preselection
requirement, the dilepton transverse momentum ($\pt^{\ell\ell}$) is required to be
larger than 50\GeV to reject the bulk of DY background events.

Since only a small amount of hadronic activity is expected in the final state of both DM and
unparticle events, any event having two or more jets with $\pt>30\GeV$ is rejected.
Processes involving top quarks are further suppressed with the use of techniques based on soft-muon and secondary-vertex $\cPqb$ jet tagging,
aimed at identifying the $\cPqb$ quarks produced in top quark decays.
Soft muons are identified using a specialised low-$\pt$ set of identification criteria focused on the muon candidate track
quality. The rejection of events with
soft muons having $\pt >3\GeV$ reduces the background from semileptonic decays of B mesons.
The $\cPqb$ jet tagging technique employed is based
on the ``combined secondary vertex'' algorithm~\cite{Chatrchyan:2012jua,CMS-PAS-BTV-15-001}.
The algorithm is calibrated to provide, on average, 80\% efficiency for tagging jets originating from $\cPqb$ quarks,
and 10\% probability of light-flavor jet misidentification.
Events are rejected if at least one \cPqb-tagged jet is reconstructed with $\pt>20\GeV$ within the tracker acceptance ($\abs{\eta} < 2.5$).

For the final selection, further kinematic requirements are set in order to achieve the best possible signal extraction. A minimal $\ETm$ of $80\GeV$ is required.
The angle between the \Z boson and the missing transverse momentum in the transverse plane $\Delta \phi_{\ell\ell,\ptvecmiss}$
is required to be larger than $2.7$ radians. The momentum balance of the event defined by ${\abs{\ETm-\pt^{\ell\ell}}/\pt^{\ell\ell}}$
is required to be smaller than $0.2$.
These variables suppress background processes such as DY and top quark production.
The event selection criteria used for the electron and muon channels are the same. They are summarized in Table~\ref{tab:selectioncuts}.

\begin{figure}[htb!]
\centering
\includegraphics[width=0.48\textwidth]{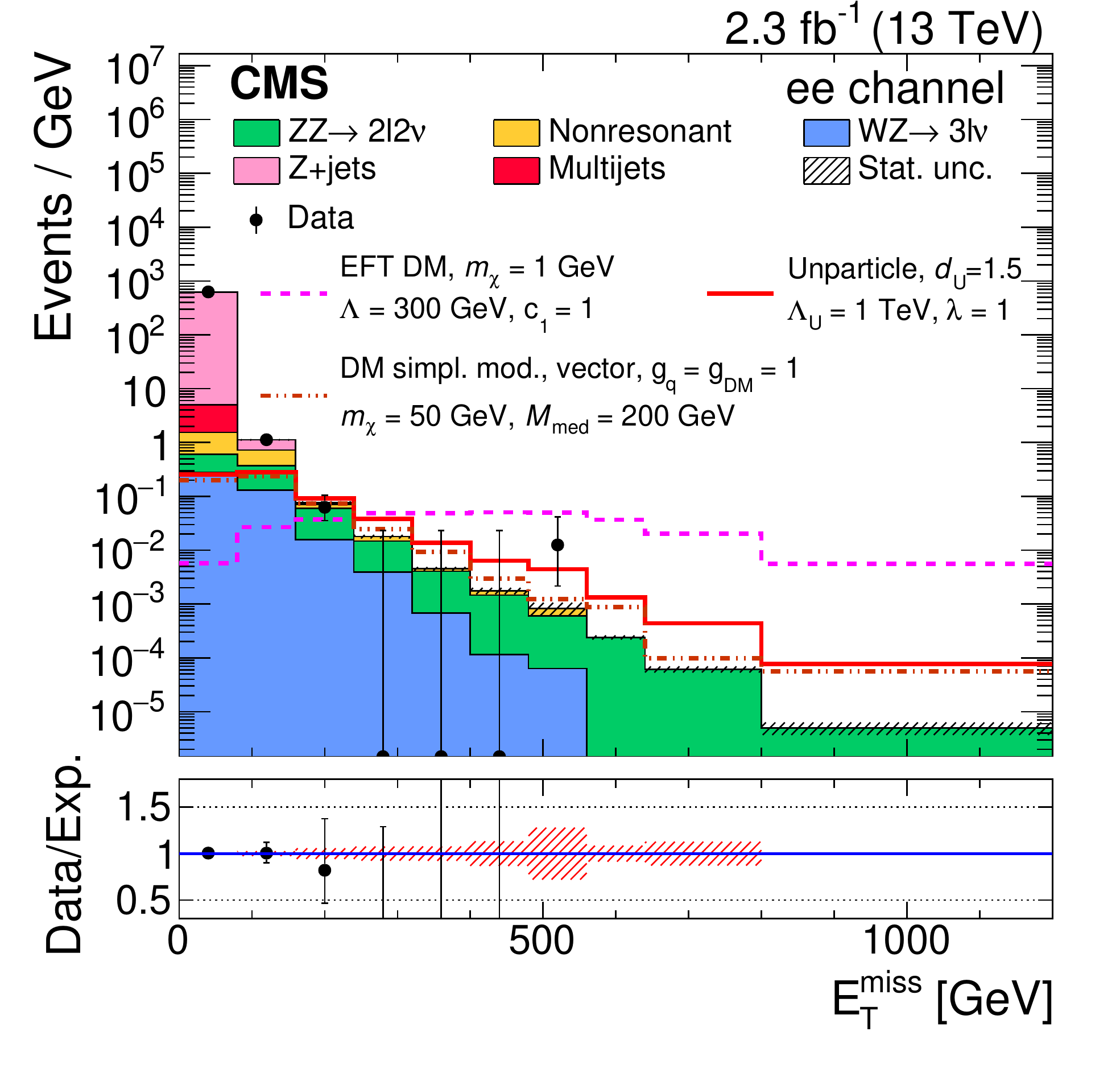}
\includegraphics[width=0.48\textwidth]{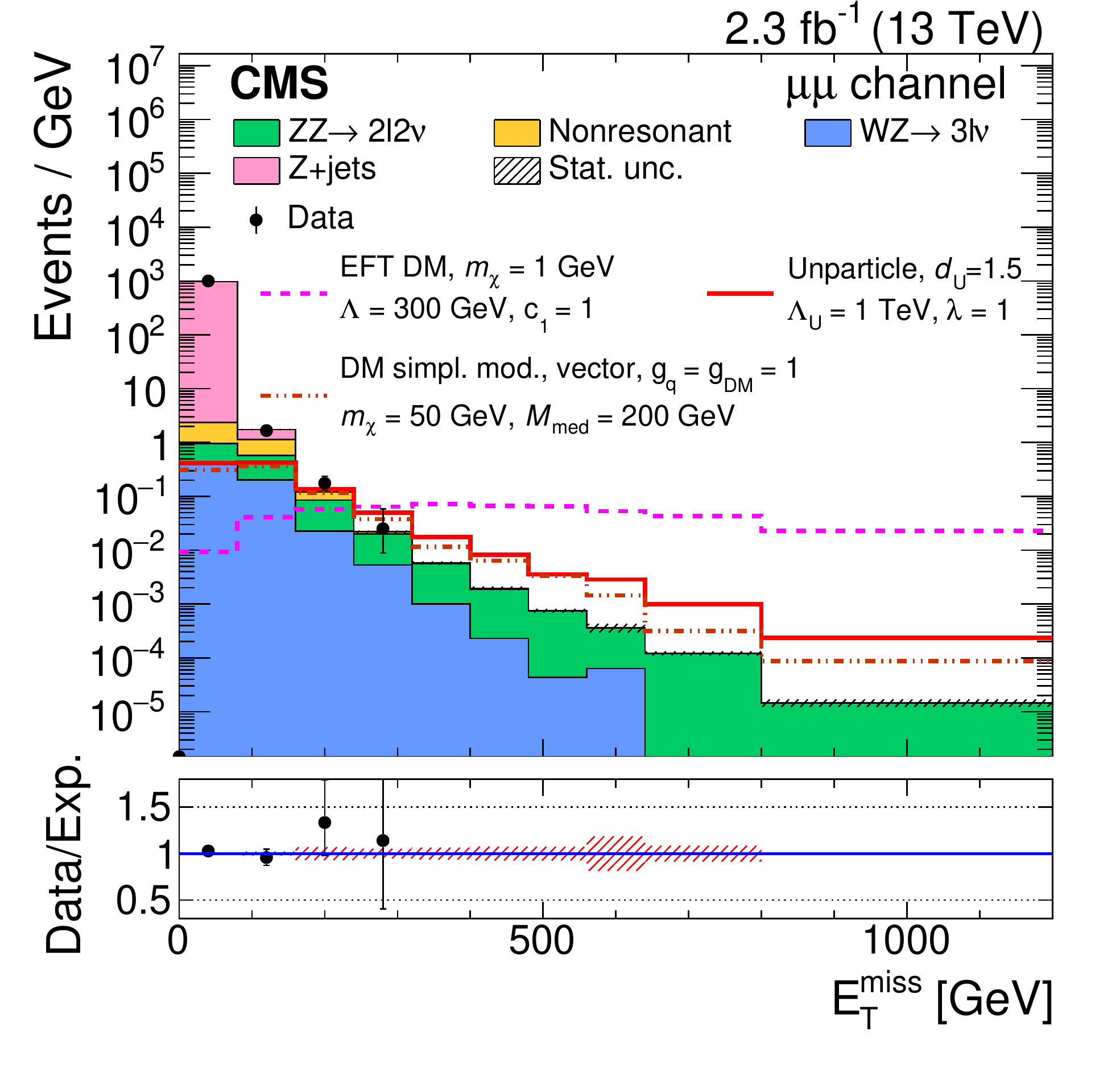}
\caption{The distribution of \ETm after preselection for the $\Z \to \Pep\Pem$ (\cmsLeft) and $\Z \to \Pgmp\Pgmm$ (\cmsRight) channels.
         Representative expected signal distributions are shown for the simplified model of DM production with vector couplings,
         the EFT scenario of DM production, and unparticles.
         The SM expectation is based on simulation only.
         The total statistical uncertainty in the overall background prediction is shown as a hatched region.
         Overflow events are included in the rightmost bins.
         The upper error bars on data points are shown for bins with zero entries (Garwood procedure) in the region up to the last non-zero entry.
         In the lower panels, the ratio between data and predicted background is shown.
        }
\label{fig:pfmet_presel}
\end{figure}

Figure~\ref{fig:pfmet_presel} shows the distributions of \ETm after preselection in the $\Pe\Pe$ and $\Pgm\Pgm$ channels.

\begin{table}[!htb]
  \centering
  \topcaption{Summary of selections used in the analysis.}
  \begin{tabular} {llll}
  \hline
                                & Variable & \multicolumn{2}{l}{Requirements}\\
  \cline{2-4}\noalign{\vskip 1mm}
\multirow{5}{*}{Preselection}   & $\pt^{\ell}$                              &   \multicolumn{2}{l}{$>$20\GeV}  \\
                                & $\abs{m_{\ell\ell} - m_{\Z}}$             &   \multicolumn{2}{l}{$<$10\GeV}  \\
                                & Jet counting                              &    \multicolumn{2}{l}{$\leq$1 jet with $\pt^\mathrm{j}>30$\GeV}  \\
                                & $\pt^{\ell\ell}$                          &    \multicolumn{2}{l}{$>$50\GeV}       \\
                                & $3^\mathrm{rd}$-lepton veto                  &    \multicolumn{2}{l}{$\pt^{\Pe,\Pgm}>10$\GeV, $\pt^{\Pgt}>20$\GeV}       \\
                                & Top quark veto                            &    \multicolumn{2}{l}{Veto on $\cPqb$ jets and soft muons} \\ \hline\noalign{\vskip 1mm}
\multirow{3}{*}{Selection}      & $\Delta \phi_{\ell\ell,\ptvecmiss}$       &    \multicolumn{2}{l}{$>$2.7 radians}   \\
                                & $\abs{\ETm-\pt^{\ell\ell}}/\pt^{\ell\ell}$    &    \multicolumn{2}{l}{$<$0.2}    \\
                                & \ETm                                      &    \multicolumn{2}{l}{$>$80\GeV}  \\
  \hline
  \end{tabular}
  \label{tab:selectioncuts}
\end{table}

\section{Background estimation}
\label{sec:backgrounds}

The $\ZZ$ and $\WZ$ backgrounds are modeled using MC simulation, and
normalized to their respective NLO cross sections. Other backgrounds, including
$\ttbar$, $\tw$, $\WW$, $\Z \to\Pgt\Pgt$, single top quark, and DY production are estimated from data for the final selection.

The simulation of the \ZZ\ process includes the $\PQq\PAQq$- and $\Pg\Pg$-induced production modes.
In order to correct the \ZZ\ differential cross section from NLO to NNLO in QCD,
$\Delta\phi(\cPZ,\cPZ)$-dependent $K$-factors are applied~\cite{Grazzini:2015xyz}.
We apply NLO electroweak (EW) $K$-factors as a function of the \pt of the
trailing boson, following the calculations in
Refs.~\cite{Bierweiler:2013dja,Gieseke:2014gka,PhysRevD.88.113005}. Electroweak
corrections to $\PW\cPZ$ production are also available, but considered small~\cite{PhysRevD.88.113005} and
not applied.

The background processes involving $\Pe\Pe$ or $\Pgm\Pgm$ pairs not directly resulting from the decay of a \Z boson are referred to as nonresonant
backgrounds. These backgrounds arise mainly from leptonic $\W$ boson decays in $\ttbar$, $\tw$,
and $\WW$ events.
There are also small contributions from the $s$- and $t$-channel single top quark events, $\Wjets$ events,
and $\Z \to\Pgt\Pgt$ events in which $\Pgt$ lepton decays result in electrons or muons and \ETm.
We estimate these nonresonant backgrounds using a data control sample, consisting of events with
opposite-charge different-flavor dilepton pairs ($\Pe^{\pm}\Pgm^{\mp}$) that otherwise pass
the full selection.
As the decay rates for $\Z \to\Pep\Pem$ and $\Z \to\Pgmp\Pgmm$ are almost equal, by equating
the ratio of observed dilepton counts to the square of the ratio of
efficiencies,
the nonresonant backgrounds in the $\Pe\Pe$ and $\Pgm\Pgm$ channels can be estimated from the $\Pe\Pgm$ channel:
\begin{linenomath*}
\begin{equation*}
\begin{aligned}
N^{\text{est}}_{\text{bkg},\Pe\Pe} &= N^{\text{data, corr}}_{\Pe\Pgm} \, k_{\Pe\Pe}, &
k_{\Pe\Pe} &= \frac12 \sqrt{\frac{N^{\text{data}}_{\Pe\Pe}}{N^{\text{data}}_{\Pgm\Pgm}}},\\
N^{\text{est}}_{\text{bkg},\Pgm\Pgm} &= N^{\text{data, corr}}_{\Pe\Pgm} \, k_{\Pgm\Pgm}, &
k_{\Pgm\Pgm} &= \frac12 \sqrt{\frac{N^{\text{data}}_{\Pgm\Pgm}}{N^{\text{data}}_{\Pe\Pe}}},
\end{aligned}
\end{equation*}
\end{linenomath*}
in which the coefficient of $1/2$ in the transfer factors $k_{\Pe\Pe}$ and $k_{\Pgm\Pgm}$
comes from the dilepton decay ratios for $\Pe\Pe$, $\Pgm\Pgm$, and $\Pe\Pgm$ in these nonresonant backgrounds,
and $N^\text{data}_{\Pe\Pe}$ and $N^\text{data}_{\Pgm\Pgm}$ are the numbers of selected $\Pe\Pe$ and $\Pgm\Pgm$ events
from data with masses in the $\Z$ boson mass window.
The ratio $\sqrt{\smash[b]{N^\text{data}_{\Pe\Pe}/N^\text{data}_{\Pgm\Pgm}}}$ and
the reciprocal quantity take into account the difference between the electron and muon selection efficiencies.
The term $N^\text{data, corr}_{\Pe\Pgm}$ is the number of $\Pe\Pgm$ events
observed in data corrected by subtracting the estimated $\ZZ$, $\WZ$, and DY background contributions.
The kinematic distributions of the estimated nonresonant backgrounds are
obtained from simulation with the overall normalization determined by the method described above.
The validity of this procedure for predicting nonresonant backgrounds is checked with simulated events containing
$\ttbar$, $\tw$, $\WW$, $\Wjets$, and $\Z \to\Pgt\Pgt$ processes.
We assign a systematic uncertainty of 26\% for this background estimation
in both the electron and muon channels for $\ETm>80$\GeV, based on closure tests that compare the predictions obtained from the control sample with those from the simulated events.

The DY process is dominant in the region of low \ETm.
This process does not produce undetectable particles, and therefore the measured \ETm arises from
limited detector acceptance and mismeasurement of particle momenta.
The estimation of this background uses simulated DY events, which are normalized to data with scale factors
obtained by measuring the number of DY events in a background-dominated control region,
after subtracting other processes. These scale factors are of order 1.0--1.2.
The control region is defined by applying the full selection with the \ETm requirement inverted.
The reliability of this approach in the high-\ETm regime has been studied by considering variables sensitive to \ETm mismeasurement, such as
the angular separation between the \ETm direction and any jet. A normalization uncertainty of 100\%, which accommodates any differences observed in these control regions, is assigned for the DY background estimate.
The assigned uncertainty has little impact on the overall signal sensitivity because of the small overall contribution from the DY background prediction.

Contributions from QCD production of multijet events is estimated using simulation and found to be negligible after final selection.

\section{Efficiencies and systematic uncertainties}
\label{sec:systematics}

The efficiencies for selecting, reconstructing, and identifying isolated leptons are determined
from simulation, and corrected with scale factors determined from applying a ``tag-and-probe'' technique~\cite{CMS:2011aa} to $\Z \to\ell^+\ell^-$ events in data.
The trigger efficiencies for the electron and muon
channels are found to be above 90\%, varying as a function of $\pt$ and $\eta$ of the lepton. The identification
efficiency, when applying the selection criteria described in Section~\ref{sec:object}, is found to be about 80--86\% for electrons and 95\% for muons,
depending on the $\pt$ and $\eta$ of the corresponding lepton.
The corresponding data-to-MC scale factors are typically in the range 0.96--1.00 for the electron and 0.96--0.98 for the muon channel,
depending on the $\pt$ and $\abs{\eta}$ of the lepton candidate.
The lepton momentum scale uncertainty is computed by varying the momentum of the leptons by its uncertainties. The lepton momentum
uncertainty is 1\% for the muons, while the uncertainty for the electrons is 2\% in the barrel and 5\% in the endcaps.
For both channels, the overall uncertainty in the efficiency of selecting and reconstructing leptons in an event is about 3\%.

In the treatment of systematic uncertainties, both normalization effects, which only affect the overall size of individual contributions,
as well as shape uncertainties, which also affect their distribution, are taken into account. The systematic
uncertainties are summarized in Table~\ref{tab:syst}. Where applicable, the symbol V is used to refer to both \Z and \W bosons.
The impact of each source of uncertainty on the observed strength of a potential signal is also reported.
The signal strength is defined as the ratio of the observed or excluded signal cross-section to the signal cross-section predicted by theory.
To calculate the impact, a maximum likelihood fit of the combined background and signal model to the expected distribution
for unity signal strength is performed. The fit is repeated with each individual nuisance parameter varied by its uncertainty.
The impact of the uncertainty is then defined as the relative change induced in the expected best fit signal strength
by the variation of the respective parameter. In the table, the reference signal is the simplified model DM scenario
with a vector mediator of mass 200\GeV, a DM particle mass of 50\GeV, and coupling $g_{\Pq}=1.0$.

\begin{table*}[htb]
\centering
\topcaption{Summary of systematic uncertainties. Each background
  uncertainty represents the variation of the relative yields of the
  particular background components. The signal uncertainties represent
  the relative variations in the signal acceptance, and the ranges quoted
  cover both signals of DM and unparticles with different DM masses or
  scaling dimensions. For shape uncertainties, the numbers correspond
  to the overall effect of the shape variation on the yield or
  acceptance. The symbol ``\NA''\  indicates that the systematic
  uncertainty is not applicable. The impact of each group of
  systematic uncertainties is calculated by performing a maximum likelihood fit
  to obtain the signal strength with each parameter separately varied by
  its uncertainty. The number given in the impact column is the relative change of the expected
  best fit signal strength that is introduced by the variation for the simplified model
  signal scenario with a vector mediator of mass 200\GeV, DM of mass 50\GeV, and coupling $g_{\Pq}=1.0$.}

\begin{tabular}{lccc}
\hline
 \multirow{2}{*}{Source of uncertainty} & Background       & Signal           & \multirow{2}{*}{Impact (\%)}   \\
                                        & uncertainty (\%) & uncertainty (\%) &                       \\
\hline
Integrated luminosity                   & 2.7              & 2.7              & 5                       \\
Lepton trigger \& identification efficiency & 3--4           & 3--4             & 2--4                     \\
Lepton momentum scale, resolution       & 1--7              & $<$1             & 1--2                     \\
Jet energy scale, resolution            & 0.1--4.0            & $<$1             & 2                       \\
$\cPqb$ jet tagging efficiency          & $<$1             & $<$1             & $<$1                     \\
Pileup                                  & 1-2              & 0.5--1.0            & 2                       \\\hline
PDF, $\alpha_\mathrm{S}$               & 2-3              & $<$1             & $<$1                       \\
Factorization, renormalization scales (signal) & \NA       & 1--2              & $<$1                    \\
Factorization, renormalization scales (VV) 	   & 3--4       & \NA              & 3                       \\
Factorization, renormalization scales (VVV)    & 12        & \NA              & $<$1                     \\
EW correction for ${\qqbar\to\ZZ}$ & 5     & \NA              & 4                       \\
EW uncertainty for WZ		    & 3     		   & \NA              & $<$1                       \\\hline
DY normalization                        & 100              & \NA              & 5                       \\
$\ttbar$, $\tw$, $\WW$ normalization    & 26               & \NA              & 2--4                     \\
MC sample size (signal)                  & \NA              & 1.5-10.0           & $<$1     \\
MC sample size (\ZZ, \WZ)           		& 0.3--3.0             & \NA              & $<$1                        \\
MC sample size (DY)                      & 13            & \NA              & $<$1                        \\
MC sample size ($\ttbar$, $\tw$, $\WW$)  & 8--10             & \NA              & $<$1                        \\
\hline
\end{tabular}
\label{tab:syst}
\end{table*}

The normalization uncertainties in the background estimates from data have been described
in Section~\ref{sec:backgrounds}.
The PDF and $\alpha_{S}$ uncertainties (referred to as PDF$+\alpha_{S}$ in the following) for signal and background processes are estimated from the
standard deviation (s.d.) of weights according to the replicas provided in the NNPDF3.0 parton distribution set~\cite{Butterworth:2015oua}.
While the influence on the estimated signal acceptance arising from theory-related uncertainties is included in the limit calculation, the corresponding effect on the normalization of the signal process is not.
For the simplified model of DM production, the effect of the signal normalization uncertainty is treated separately from the experimental uncertainty and is shown as a dashed band around the observed limit.
Since the EFT benchmark and unparticle scenarios are extremely simplified, theory-related cross-section uncertainties are not considered to be realistic for these models and are thus neglected.
The efficiencies for signal, $\ZZ$, and $\WZ$ processes are estimated using simulation, and the uncertainties
in the corresponding yields are derived by varying the renormalization and factorization scales, $\alpha_{S}$, and choice of PDFs. The factorization and renormalization scale uncertainties are assessed by varying the original scales by factors of 0.5 or 2.0, and amount to 2--3\% for $\ZZ$ and $\WZ$ processes.
The effect of variations in $\alpha_{S}$ and choice of PDFs is 2\% for the $\ZZ$ and $\WZ$ backgrounds.
A 3\% normalization uncertainty is assigned to the $\WZ$ background to account for higher-order EW corrections~\cite{PhysRevD.88.113005}.
The uncertainty assigned to the integrated luminosity measurement is 2.7\%~\cite{CMS-PAS-LUM-15-001}.

Experimental sources of shape uncertainty are the lepton momentum scale, the jet energy scale and resolution, the $\cPqb$ tagging efficiency, and the pileup modeling.
The effect of each uncertainty is estimated by varying the respective variable of interest by its
uncertainties, and propagating the variations to the distribution of $\ETm$ after the final selection.
In the case of the lepton momentum scale, the uncertainty is computed by varying the momentum of the leptons by their uncertainties.
The uncertainty due to the lepton momentum scale is evaluated to be less than 1\% (1--7\%) for signal (background).

The uncertainties in the calibration of the jet energy scale and resolution directly affect
the assignments of jets to jet categories,
the \ETm computation, and all the selections related to jets. The
effect of the jet energy scale uncertainty is estimated by varying the energy scale by $\pm1$ s.d.
A similar strategy is used to evaluate the systematic uncertainty related to the jet energy resolution.
The effect of the shifts is propagated to \ETm.
The uncertainties in the final yields are found to be less than 1\% for signal and less than 4\% for background.

In order to reproduce b tagging efficiencies observed in data, an event-by-event reweighting using data-to-simulation scale factors
is applied to simulated events.
The uncertainty associated with this procedure is obtained by varying the event-by-event weight by ${\pm}1$ s.d.
The total uncertainty in the final yields due to b tagging is less than 1\% for both signal and background.
All simulated events are reweighted to reproduce the pileup conditions observed in data.
To compute the uncertainty related to pileup modeling, we shift the mean of the distribution in simulation by 5\%~\cite{Aaboud:2016mmw}.
The variation of the final yields induced by this procedure is 0.5--1\% for signal and 1--2\% for background.
For the processes estimated from simulation, the sizes of the MC samples limit the precision of the modeling, and
the resulting statistical uncertainty is incorporated into the shape uncertainty. A similar treatment is applied to the backgrounds
estimated from control samples in data, based on the statistical uncertainties in the corresponding control samples.

\section{Results}
\label{sec:results}

\begin{figure}[!htb]
\centering
\includegraphics[width=0.48\textwidth]{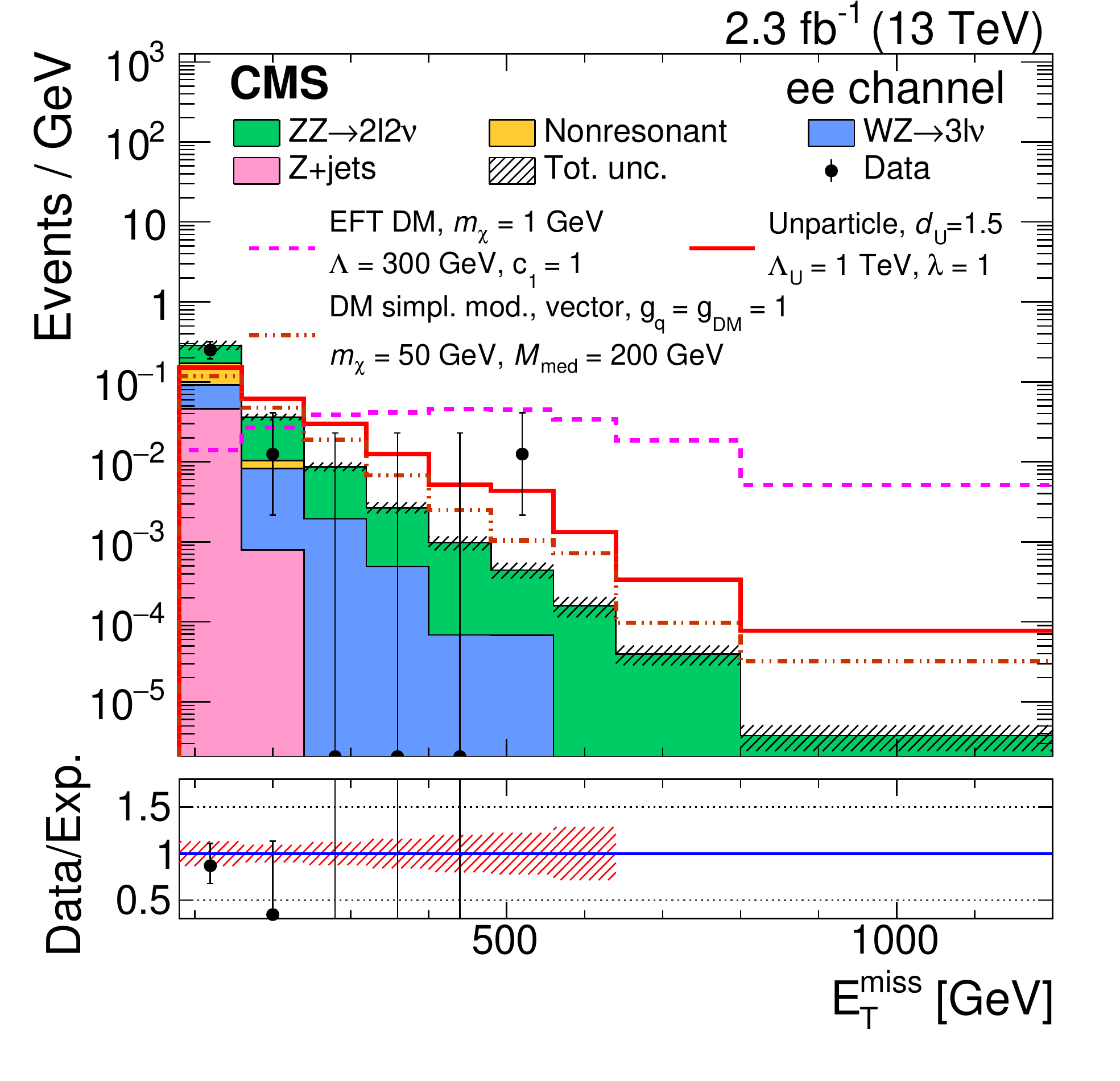}
\includegraphics[width=0.48\textwidth]{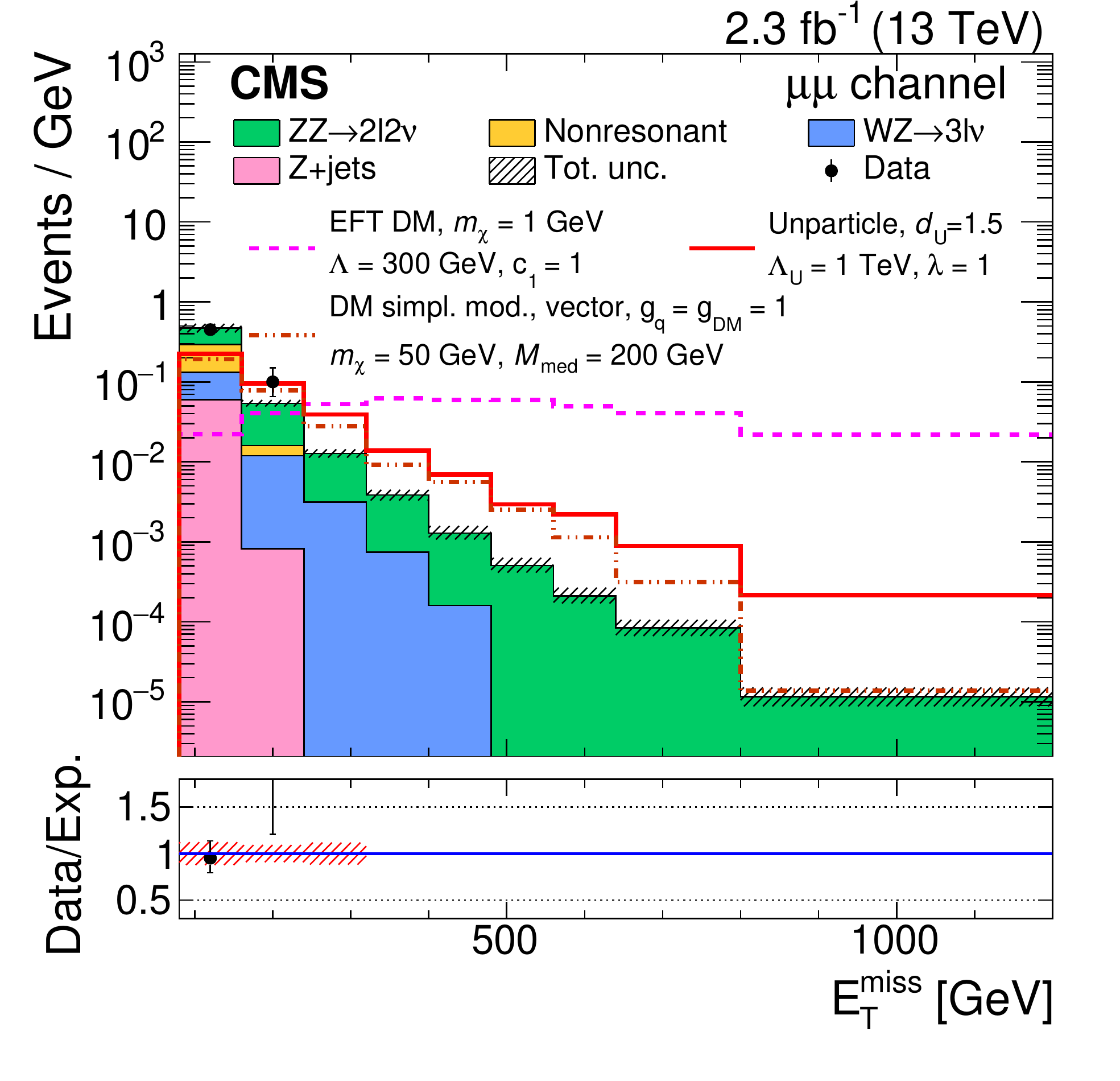}
\caption{Distributions of \ETm
        for the final selection in the $\Pep\Pem$ (left) and $\Pgmp\Pgmm$ (right) channels.
        Expected signal distributions are shown for the simplified model of DM production with vector couplings,
        the EFT DM production benchmark, and unparticle model.
        The total uncertainty (stat.~$\oplus$~sys.) in the overall background is shown as a hatched region.
        Overflow events are included in the rightmost bins.
		In the lower panels, the ratio between data and predicted background is shown.
        }
\label{fig:postfit_shapes_LogY}
\end{figure}

For both the electron and the muon channels, a shape-based analysis is employed.
The expected numbers of background and signal events scaled by a signal strength modifier
are combined in a binned likelihood for each bin of the $\ETm$ distribution.
The numbers of observed and expected events are shown in Table~\ref{tab:seltable},
which also includes the expectation for a selected parameter point for each type of signal.
Figure~\ref{fig:postfit_shapes_LogY} shows the $\ETm$ distributions after the final selection.
The observed distributions agree with the SM background predictions and no excess of events is observed.

Upper limits on the contribution of events from new physics are computed by
using the modified frequentist approach CL$_\mathrm{s}$~\cite{junkcls,Read1,HiggsCombination}.

\begin{table}[htb!]
\centering
\topcaption{Signal predictions and background estimates for the final selection with $\ETm>80\GeV$.
        The DM signal yields from the simplified model are given for mass $m_\chi=50\GeV$
        and a mediator mass $M_\text{med} =200\GeV$ for both the vector and axial-vector coupling scenarios.
        For the EFT benchmark with DM pair coupling to gauge bosons,
        the signal yields are given for  $m_\chi=1\GeV$, cutoff scale $\Lambda=300\GeV$, and the coupling $c_1=1$.
        Yields for the unparticle model are shown for scaling dimension $d_\mathsf{U}=1.5$, and cutoff scale $\Lambda = 1\TeV$.
        The corresponding statistical and systematic uncertainties are shown, in that order.}
\begin{tabular}{lll}
\hline
Process & \multicolumn{1}{c}{$\Pep\Pem$} & \multicolumn{1}{c}{$\Pgmp\Pgmm$} \\
\hline
Simplified DM model, vector mediator        & \multirow{2}{*}{15.8   $\pm$ 0.4   $\pm$ 1.0}   & \multirow{2}{*}{25.5   $\pm$ 0.5   $\pm$ 1.8} \\
$m_\chi=50\GeV$, $M_\text{med} = 200\GeV$&&\\ [1ex]
Simplified DM model, axial-vector mediator  &  \multirow{2}{*}{12.9   $\pm$ 0.3  $\pm$ 0.9}     & \multirow{2}{*}{19.2   $\pm$ 0.4   $\pm$ 1.3} \\
$m_\chi=50\GeV$, $M_\text{med} = 200\GeV$  &  & \\\hline
EFT DM model  & \multirow{2}{*}{25.4   $\pm$ 0.4   $\pm$ 2.7}   & \multirow{2}{*}{47.7   $\pm$ 0.5   $\pm$ 5.9} \\
$m_\chi=1\GeV$,  $\Lambda = 300\GeV$   & & \\[1ex]
Unparticle model  & \multirow{2}{*}{21.5 $\pm$ 0.6 $\pm$ 0.9}   & \multirow{2}{*}{31.0 $\pm$ 0.7 $\pm$ 1.6} \\
$d_\mathsf{U}=1.5$,  $\Lambda_\mathsf{U} = 1\TeV$   & & \\
\hline
                $\dyll$                                 &  4.9  $\pm$ 0.6  $\pm$ 4.9  &  5.3  $\pm$ 0.7  $\pm$ 5.3   \\
                $\WZ\to3\ell\nu$               &  4.6  $\pm$ 0.2  $\pm$ 0.4  &  7.0  $\pm$ 0.2  $\pm$ 0.6  \\
                $\ZZ\to2\ell2\nu$              & 12.4  $\pm$ 0.1  $\pm$ 1.0  & 18.7  $\pm$ 0.1  $\pm$ 1.5  \\
                $\ttbar$/$\tw$/\WW/$\Z\to\Pgt\Pgt$      &  7.0  $\pm$ 1.0  $\pm$ 1.9  & 14.0  $\pm$ 2.1  $\pm$ 3.8   \\
                VVV, $\ZZ\to2\ell 2q,4\ell$  &  $<$0.1 &  $<$0.1 \\
\hline
                         Total background               & 28.9   $\pm$ 1.2   $\pm$ 5.4   & 45.0   $\pm$ 2.2   $\pm$ 6.8   \\
\hline
                                    Data                & \multicolumn{1}{c}{22}                             & \multicolumn{1}{c}{44} \\
\hline
\end{tabular}
\label{tab:seltable}
\end{table}

\subsection{The DM interpretation}

\begin{figure}[!hbt]
  \centering
  \includegraphics[width=0.48\textwidth]{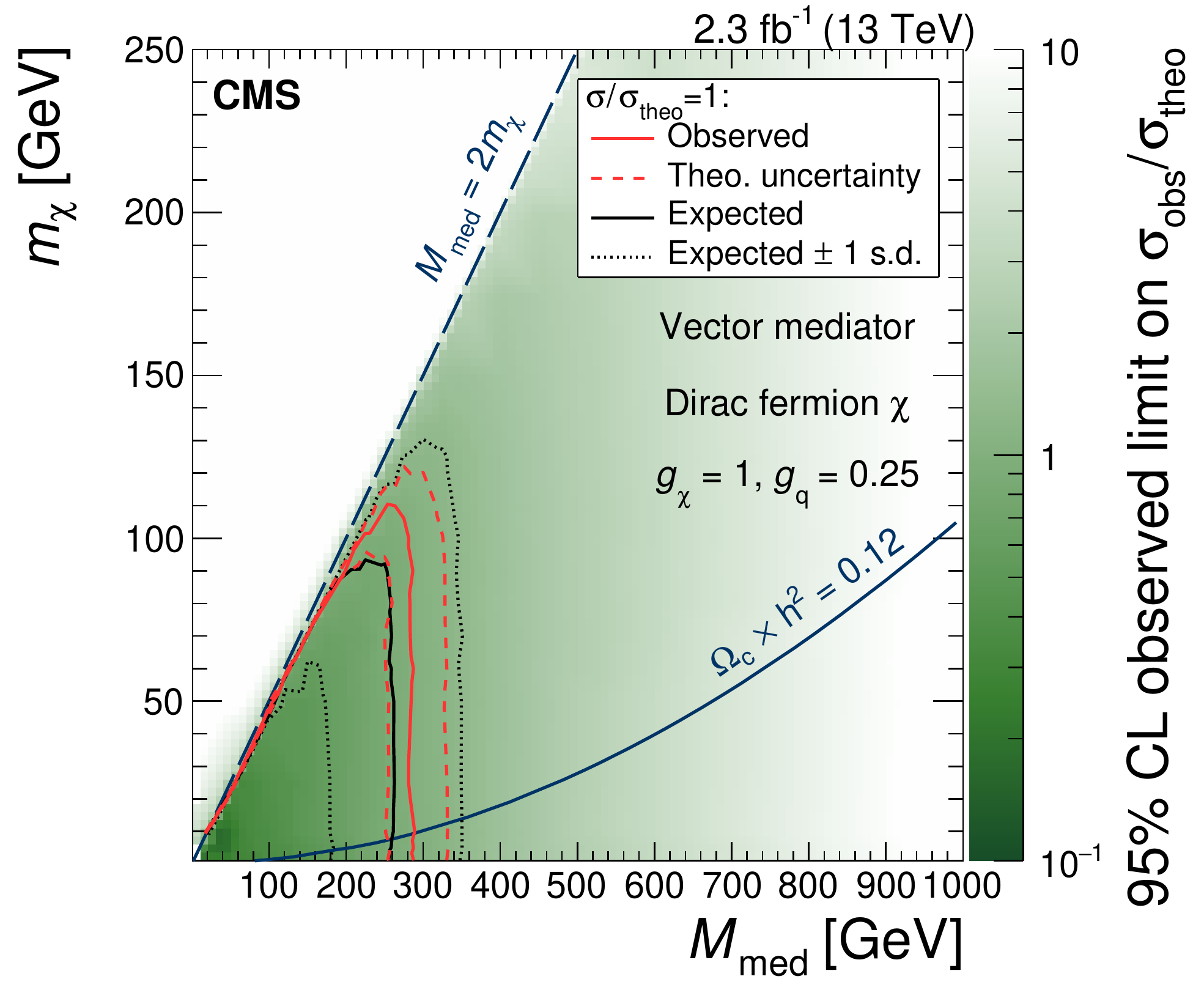}
  \includegraphics[width=0.48\textwidth]{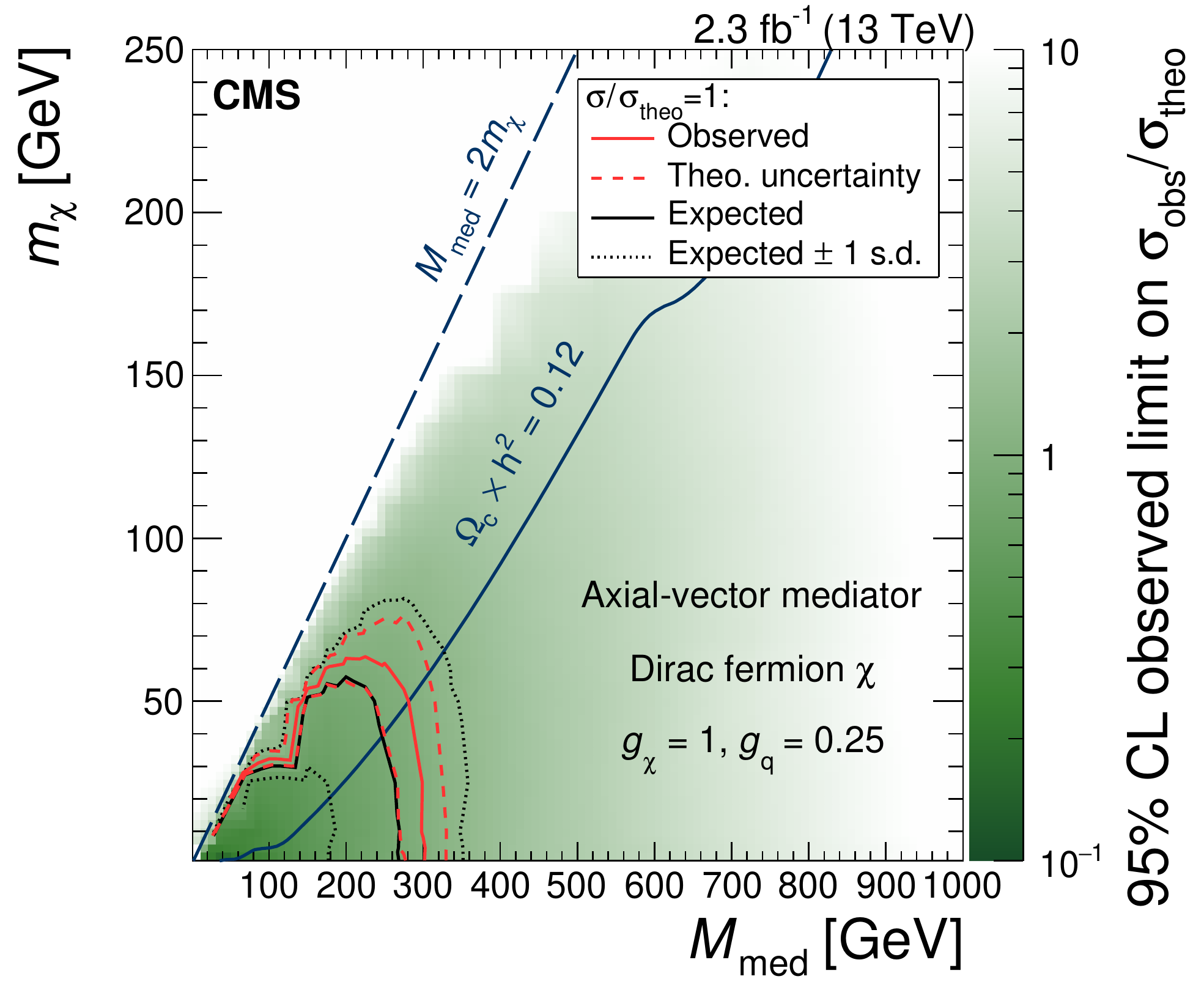}\\
  \includegraphics[width=0.48\textwidth]{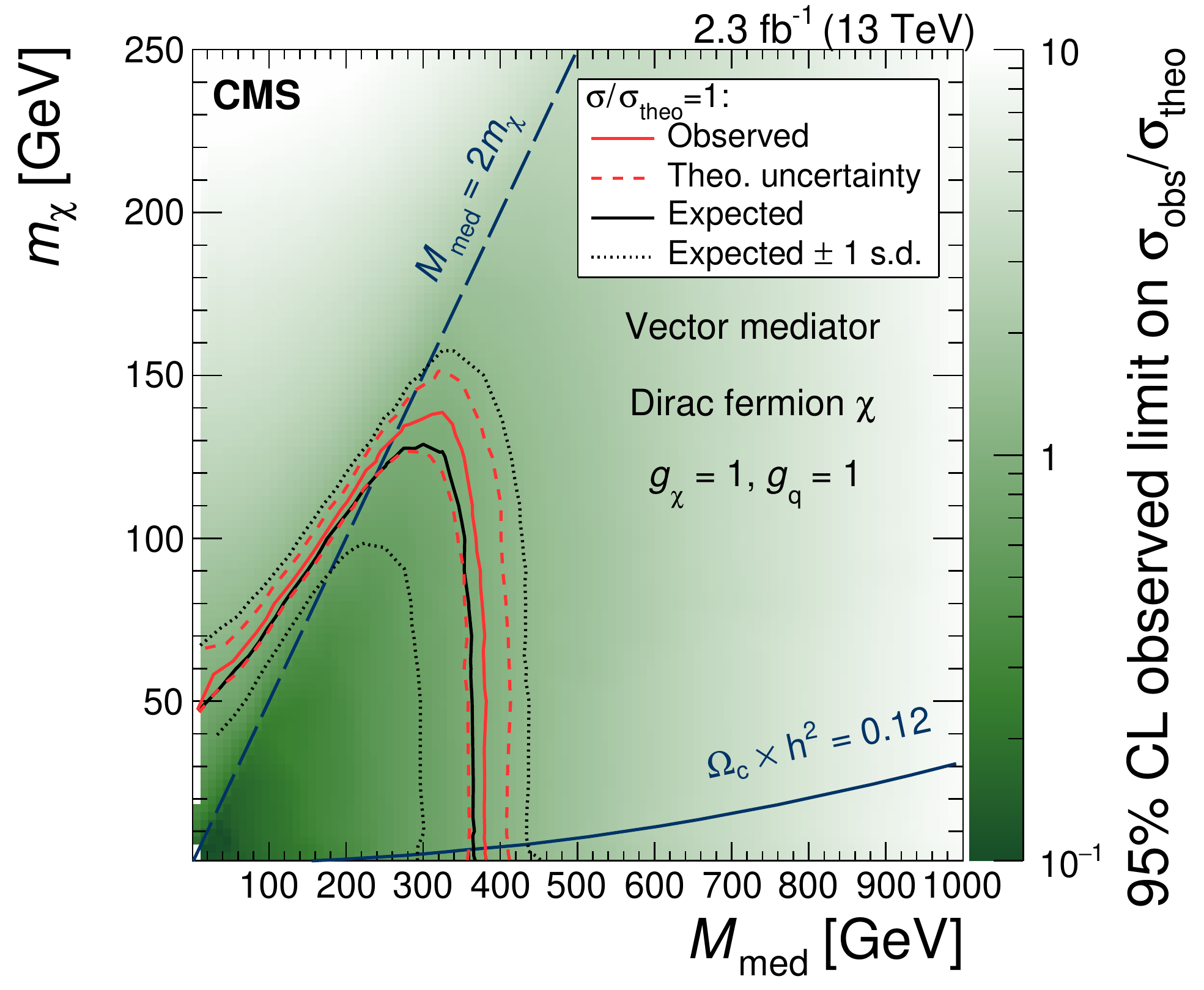}
  \includegraphics[width=0.48\textwidth]{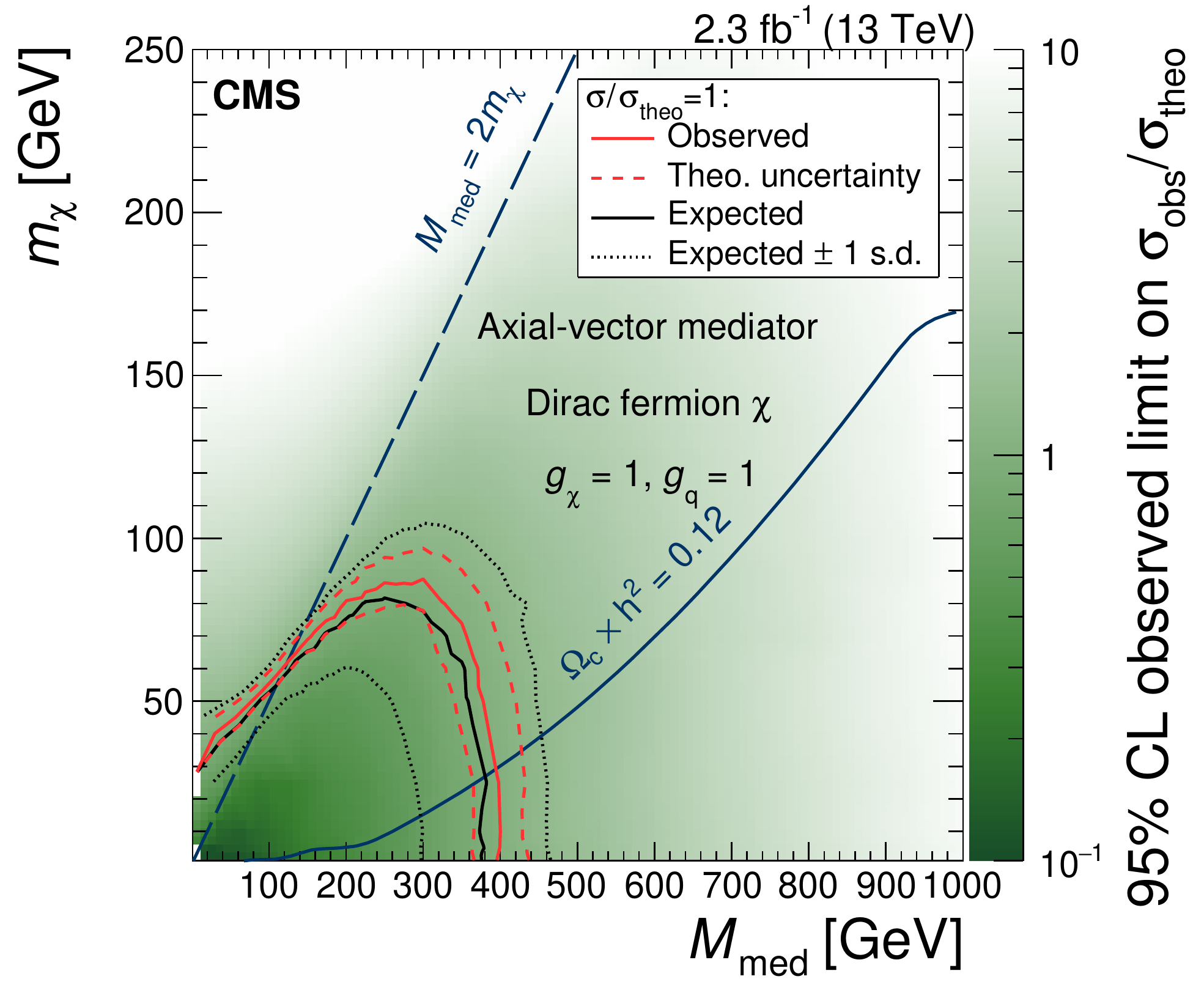}
  \caption{
			The 95\%~\CL observed limits on the signal strength $\sigma_\text{obs}/\sigma_\text{theo}$
			in both vector (left) and axial-vector (right) mediator scenarios,
			for mediator-quark coupling constant values $g_{\Pq} = 0.25$ (upper) and $1$ (lower).
			In all cases, the DM-mediator coupling $g_{\chi}$ is set to one.
			The expected exclusion curves for unity signal strength are shown as a reference, with black dashed lines indicating the expected $\pm1$ s.d. interval due to experimental uncertainties.
			The red dashed lines show the influence of theory-related signal normalization uncertainties on the observed limits, which are estimated to be 15\%.
			The solid line labeled ``$\Omega_c\times h^2 = 0.12$" identifies the parameter region where no additional new physics beyond the simplified model is necessary to reproduce the observed DM relic abundance in the universe~\cite{Boveia:2016mrp,2013ApJS..208...19H,Backovic:2013dpa,Ade:2015xua,Pree:2016hwc}.
			}
  \label{fig:DM13TeV:MV_MX}
\end{figure}

The results are interpreted in the context of a simplified model
of DM production. Figure~\ref{fig:DM13TeV:MV_MX} shows
95\%~confidence level (CL) expected and observed limits on the signal strength
$\sigma^\text{obs}/\sigma^\text{th}$
in the case of vector and axial-vector mediators and for two
possible values of the quark-mediator coupling constant, $g_{\Pq}=0.25$ or $1$.
Independent of the type of coupling, production of DM particles via an on-shell mediator ($2m_\chi<M_\text{med}$)
can be excluded up to mediator masses of $\approx$400\GeV for $g_{\Pq}=1.0$ and up to $\approx$300\GeV for $g_{\Pq}=0.25$. Dark matter particle masses are probed up to 100--150\GeV for vector and up to 50--100\GeV for axial-vector couplings. For $g_{\Pq}=1.0$, a small region of off-shell parameter space can also be excluded. In the case of $g_{\Pq}=0.25$, sensitivity is limited to the on-shell region.

The simplified model allows a calculation of the DM relic abundance in the universe for each parameter point~\cite{Pree:2016hwc,Backovic:2013dpa}.
Parameter combinations consistent with measurements of the DM relic abundance in the universe are indicated in Fig.~\ref{fig:DM13TeV:MV_MX}.
For these parameter combinations, no BSM phenomena other than the simplified model are needed to account for the relic abundance in the universe.
For other parameter values, additional phenomena, such as an extended dark sector, are necessary.

\begin{figure}[!hbt]
  \centering
  \includegraphics[width=0.48\textwidth]{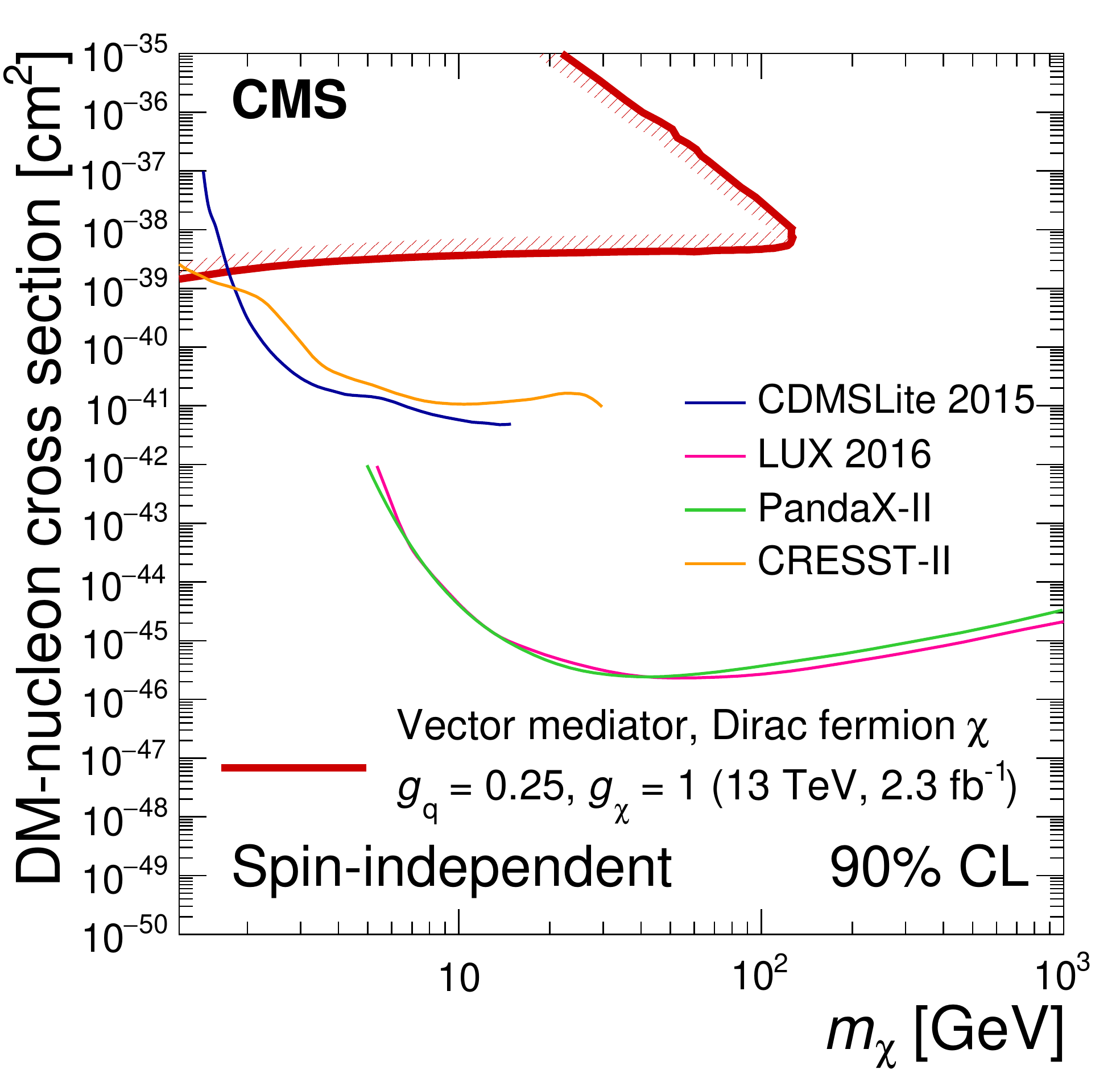}
  \includegraphics[width=0.48\textwidth]{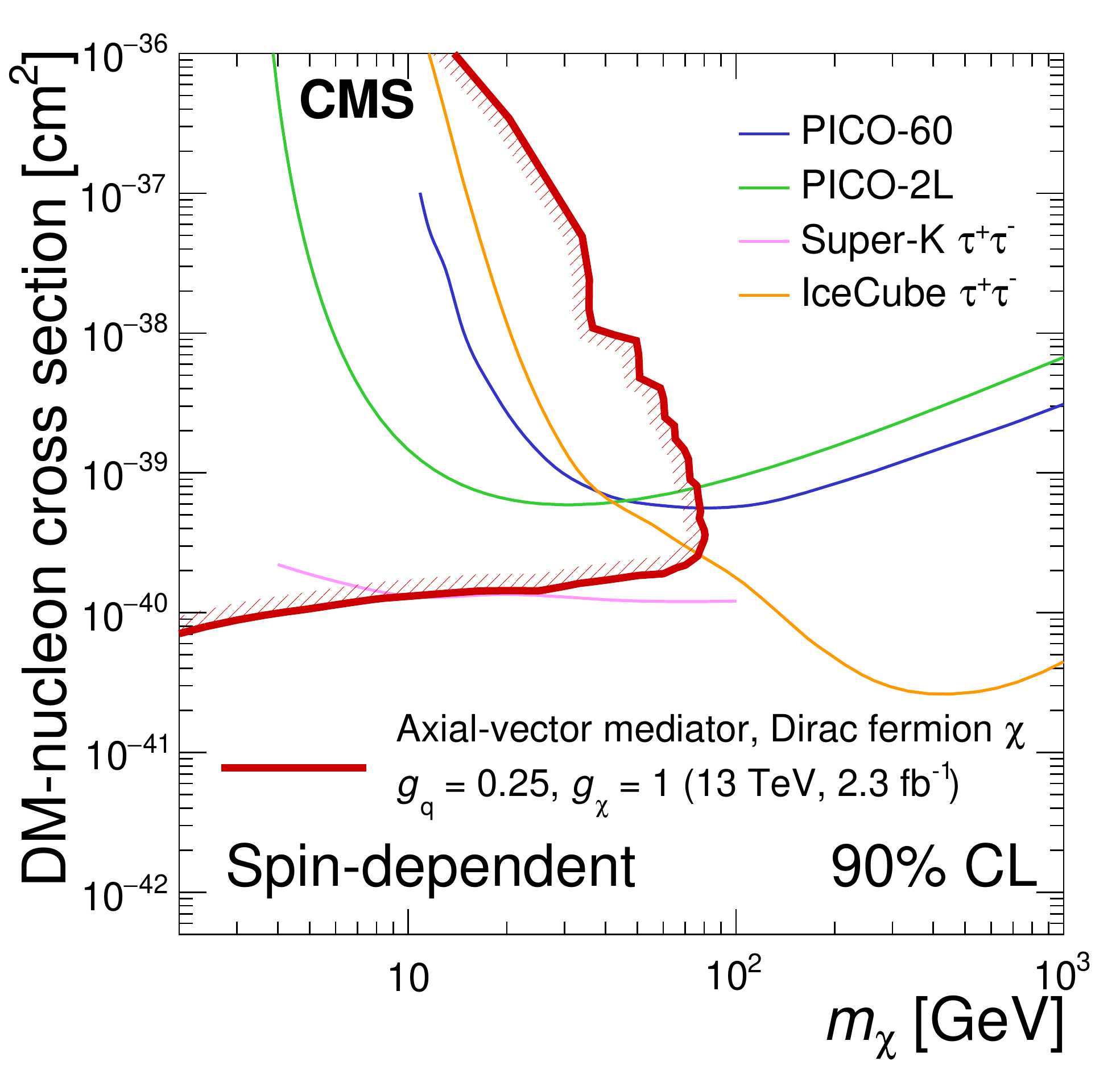}
  \caption{
			Observed 90\% CL limits on the DM-nucleon scattering cross sections
			in both spin-independent (left) and spin-dependent (right) cases,
			assuming a mediator-quark coupling constant $g_{\Pq} = 0.25$ and mediator-DM coupling constant $g_{\chi} = 1$.
			The line shading indicates the excluded region.
			Limits from the LUX~\cite{Akerib:2016vxi}, \mbox{CDMSLite}~\cite{Agnese:2015nto}, PandaX-II~\cite{Tan:2016zwf}, and CRESST-II~\cite{Angloher:2015ewa} experiments are shown for the
			spin-independent case. Limits from the Super-Kamiokande~\cite{Choi:2015ara}, PICO-2L~\cite{Amole:2016pye}, PICO-60~\cite{Amole:2015pla}, and
			IceCube~\cite{Aartsen:2016exj} experiments are shown for the spin-dependent case.
			}
  \label{fig:DM13TeV:WIMPXS}
\end{figure}

The exclusion limits in the $M_\text{med}$-$m_\chi$ plane are translated into limits on the DM-nucleon scattering cross section using the prescription of Ref.~\cite{Boveia:2016mrp}.
The limits are set at 90\%~\CL, assuming $g_{\Pq}=0.25$.
The resulting exclusion curves for both spin-independent (vector) and spin-dependent (axial-vector) cases are shown in Fig.~\ref{fig:DM13TeV:WIMPXS},
which compares them to the results from direct detection experiments. The comparison of collider and direct detection experiments highlights the complementarity of the two approaches. Especially in the case of lower DM masses and axial-vector couplings, a collider-based search can exclude parameter space not covered by direct detection experiments.
In all cases, the DM-mediator coupling $g_{\chi}$ is set to one.

\begin{figure}[!htb]
\centering
\includegraphics[width=0.48\textwidth]{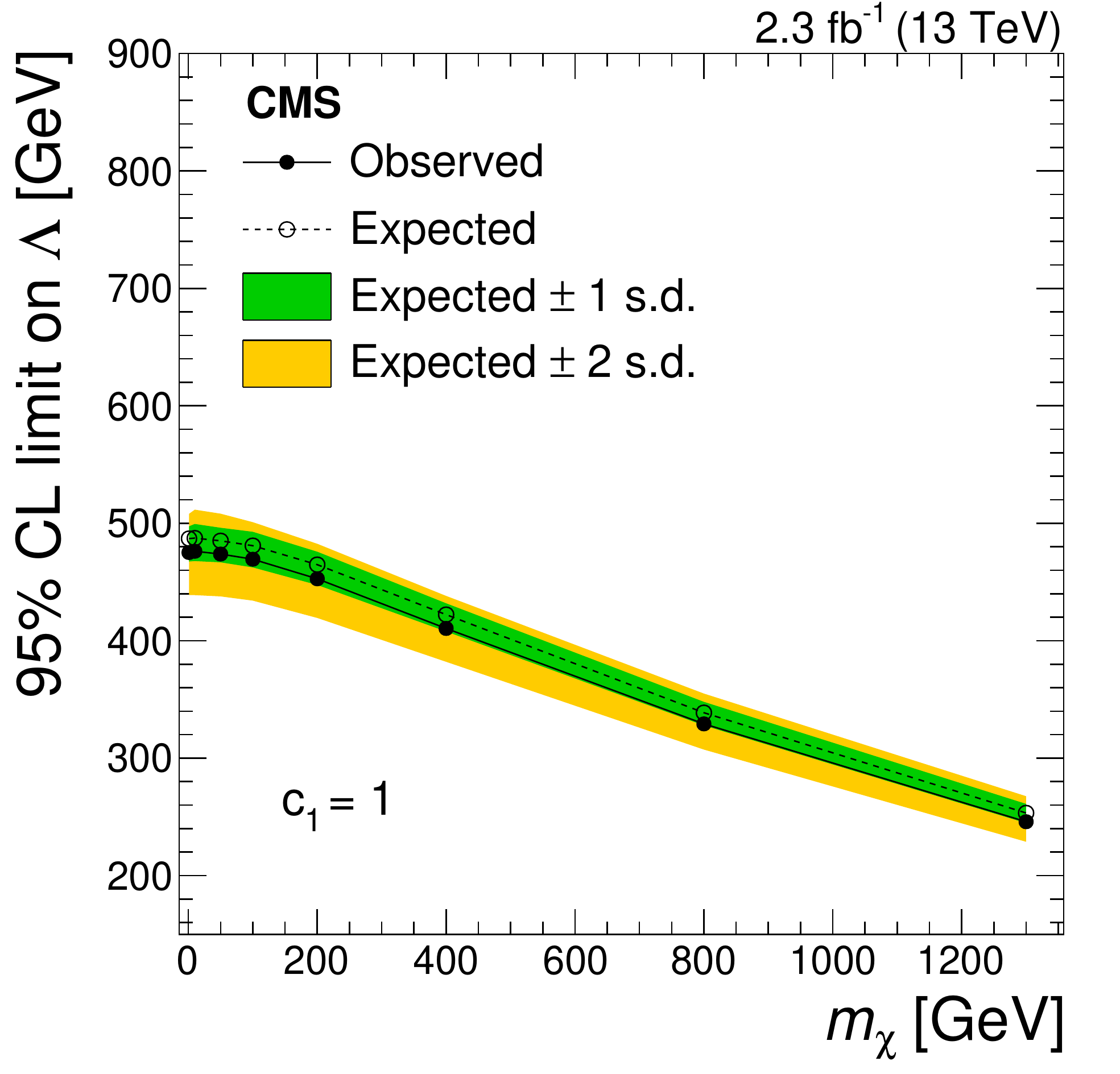}
\caption{The 95\%~\CL expected and observed limits on the cutoff scale $\Lambda$ of the EFT benchmark of DM production as a function of DM particle mass $m_\chi$.}
\label{fig:DM13TeV:EWKDM_CutoffScale_Limit}
\end{figure}

\begin{figure}[!htb]
\centering
\includegraphics[width=0.49\textwidth]{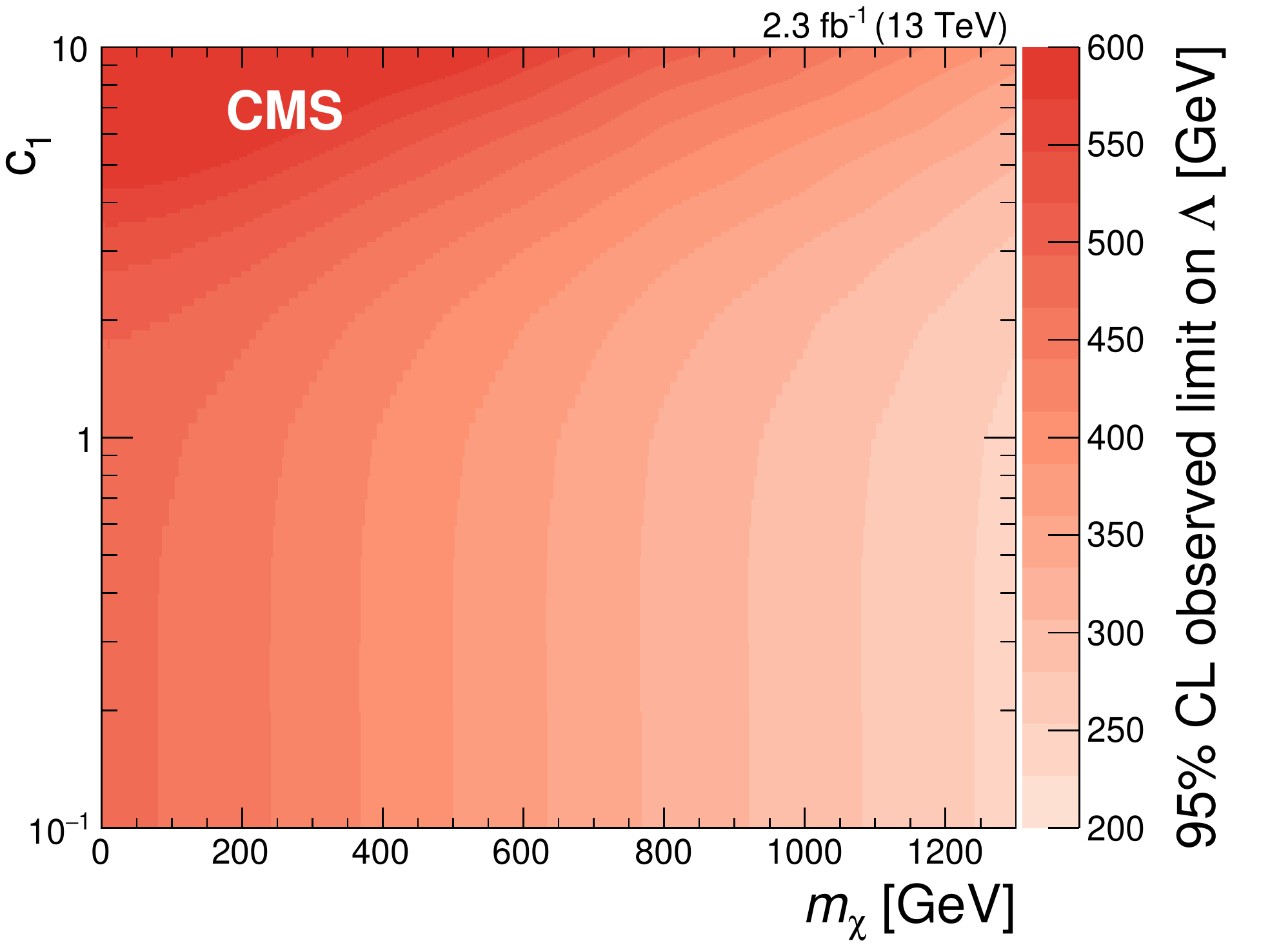}
\includegraphics[width=0.49\textwidth]{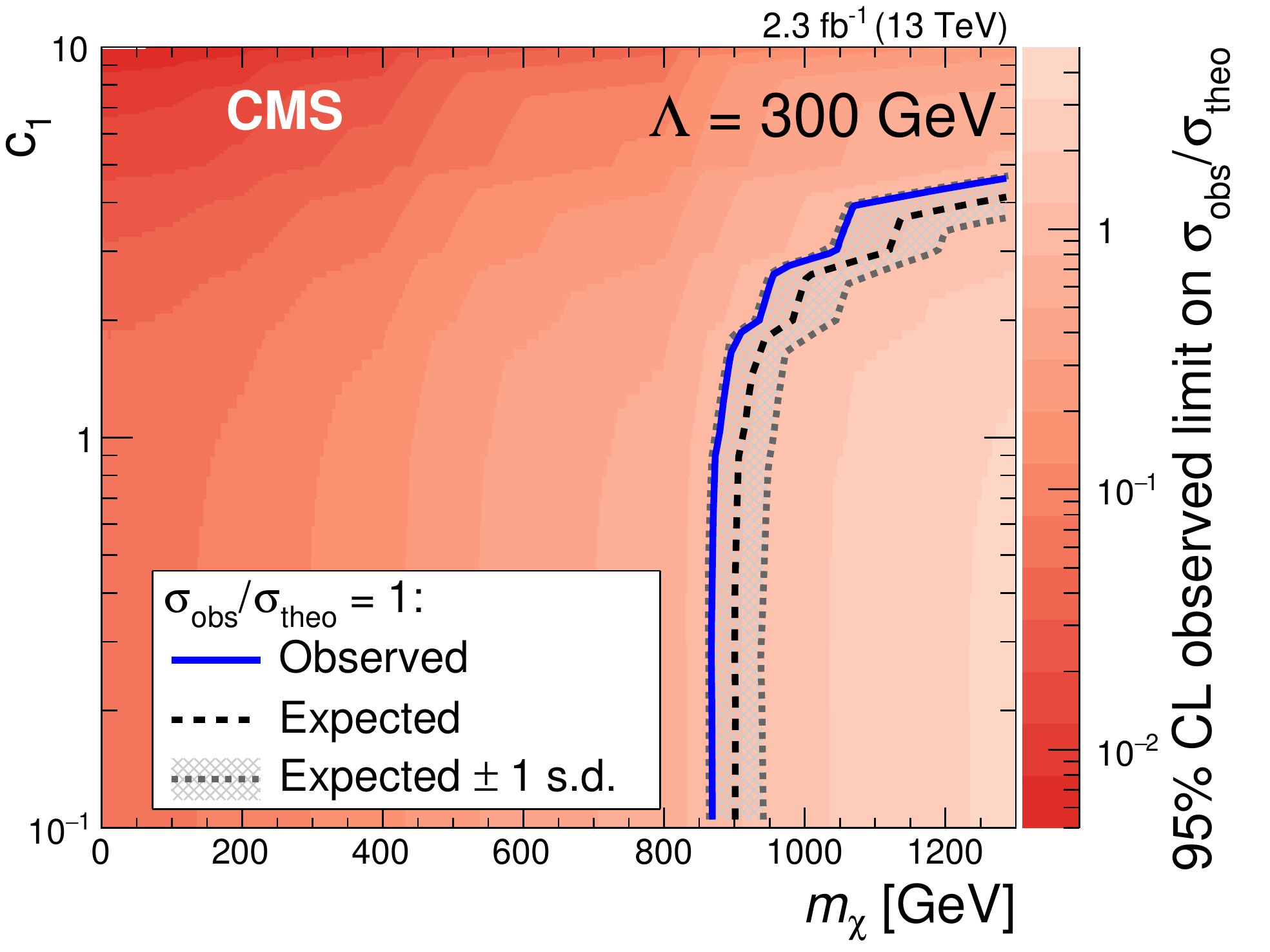}
\caption{
		The 95\%~\CL observed limits on the cutoff scale $\Lambda$ (left) and signal strength $\sigma^\text{obs}/\sigma^\text{th}$ (right)
		as a function of coupling $c_1$ and DM mass $m_\chi$. The expected exclusion curves for unit signal strength are shown as a reference.
		The gray shaded area bounded by gray dashed lines indicates the expected $\pm1$ s.d. interval due to experimental uncertainties.
		}
\label{fig:DM13TeV:EWKDM_2D_Limit}
\end{figure}

Figure~\ref{fig:DM13TeV:EWKDM_CutoffScale_Limit} shows 95\%~\CL expected limits on the cutoff scale $\Lambda$ of
the EFT benchmark model with DM pair coupling to gauge bosons. The limits are derived as a function of the DM particle mass.
At low masses, cutoff scales up to $\approx480\GeV$ can be excluded. With increasing DM particle mass, sensitivity decreases with $\Lambda<250\GeV$ excluded for $m_\chi=1.3\TeV$.
The 95\%~\CL expected limits on the cutoff scale $\Lambda$ and signal strength $\sigma^\text{obs}/\sigma^\text{th}$
as a function of coupling $c_1$ and DM mass $m_\chi$ are shown in Fig.~\ref{fig:DM13TeV:EWKDM_2D_Limit}.
At $c_1\approx 1$, the interaction is dominated by the $ZZ\chi\chi$-vertex.
With increasing $c_1$, the $\gamma\Z\chi\chi$-vertex begins to contribute, yielding an improvement in the sensitivity.

\subsection{Unparticle interpretation}

\begin{figure}[htb!]
\centering
\includegraphics[width=0.48\textwidth]{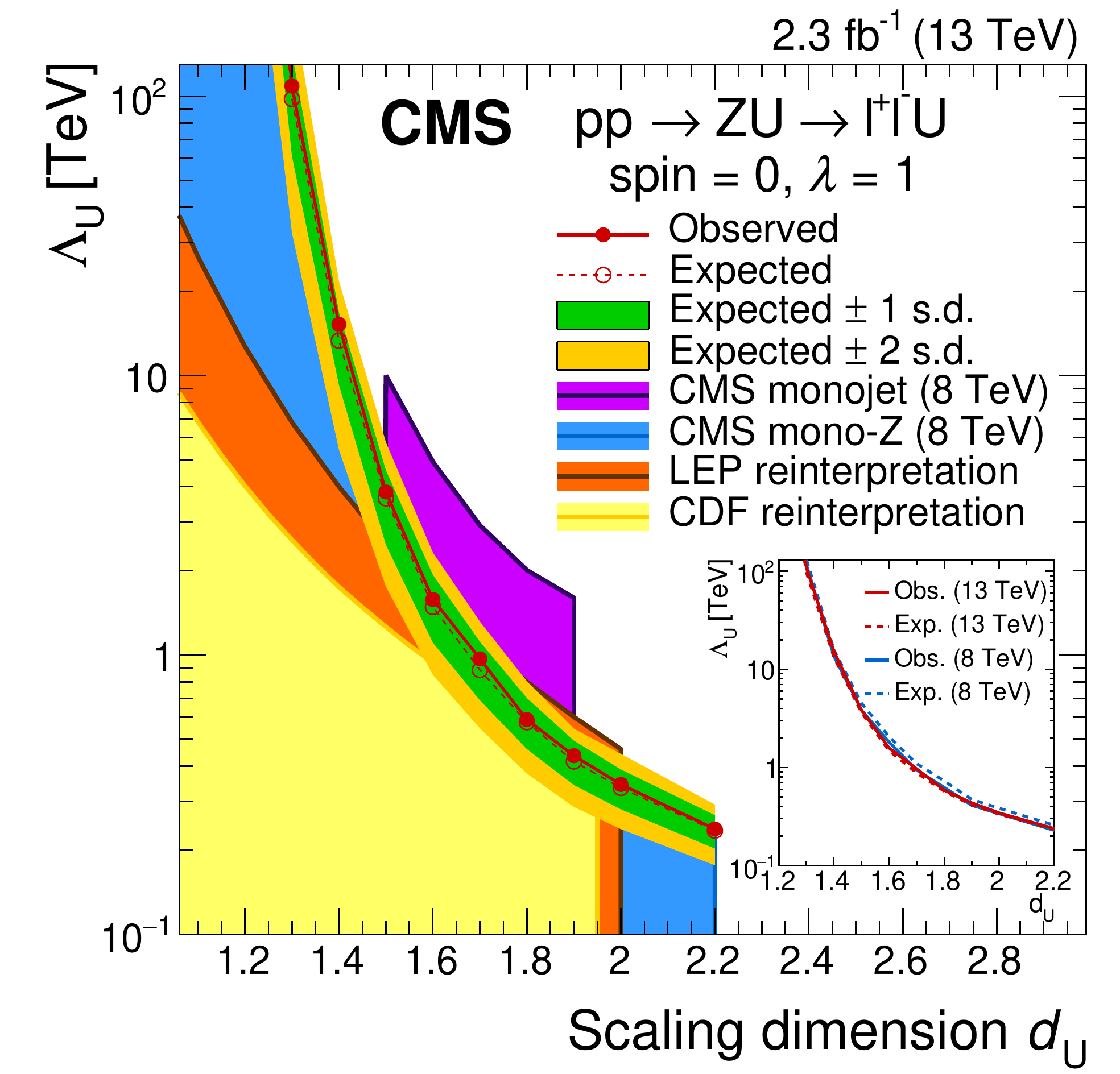}
\caption{
			The 95\% \CL lower limits on the effective unparticle cutoff scale $\Lambda_\textsf{U}$ for a fixed coupling $\lambda=1$.
			The results from the CMS monojet~\cite{Khachatryan:2014rra} and mono-Z~\cite{Khachatryan:2015bbl} searches,
			as well as a reinterpretation of LEP and CDF searches~\cite{Kathrein:2010ej} are shown for comparison.
			The LEP results assume a coupling of unparticles to \Z bosons and photons.
			The CDF (CMS) monojet result is based on a gluon-unparticle coupling operator (gluon- and quark-unparticle coupling operators).
			The inset compares the expected and observed limits for the CMS mono-Z analyses at $\sqrt{s}=8$ and 13\TeV.
			Note that the cutoff scales $\Lambda_\textsf{U}$ for different operators do not have to be identical.
			Consequently, the comparison shown here with the results other than the CMS 8\TeV mono-Z analysis is  only qualitative.
        }
\label{fig:unparticleLimits}
\end{figure}

In the unparticle scenario, 95\% CL lower limits are set on the
effective cutoff scale $\Lambda_{\textsf U}$. A fixed coupling $\lambda=1$ is assumed.
The limits on $\Lambda_{\textsf U}$ are shown in Fig.~\ref{fig:unparticleLimits} as a function of the scaling dimension $d_{\textsf U}$.
The result is compared with the limits obtained from previous CMS searches in the monojet~\cite{Khachatryan:2014rra}
and mono-Z~\cite{Khachatryan:2015bbl} channels, as well as with a reinterpretation of LEP searches~\cite{Kathrein:2010ej}.
Comparable sensitivity to the previous CMS mono-Z search is achieved owing to the increase in collision energy, which offsets the larger size of the previous dataset.

\subsection{Model-independent limits}
\begin{figure}[htb!]
\centering
\includegraphics[width=0.48\textwidth]{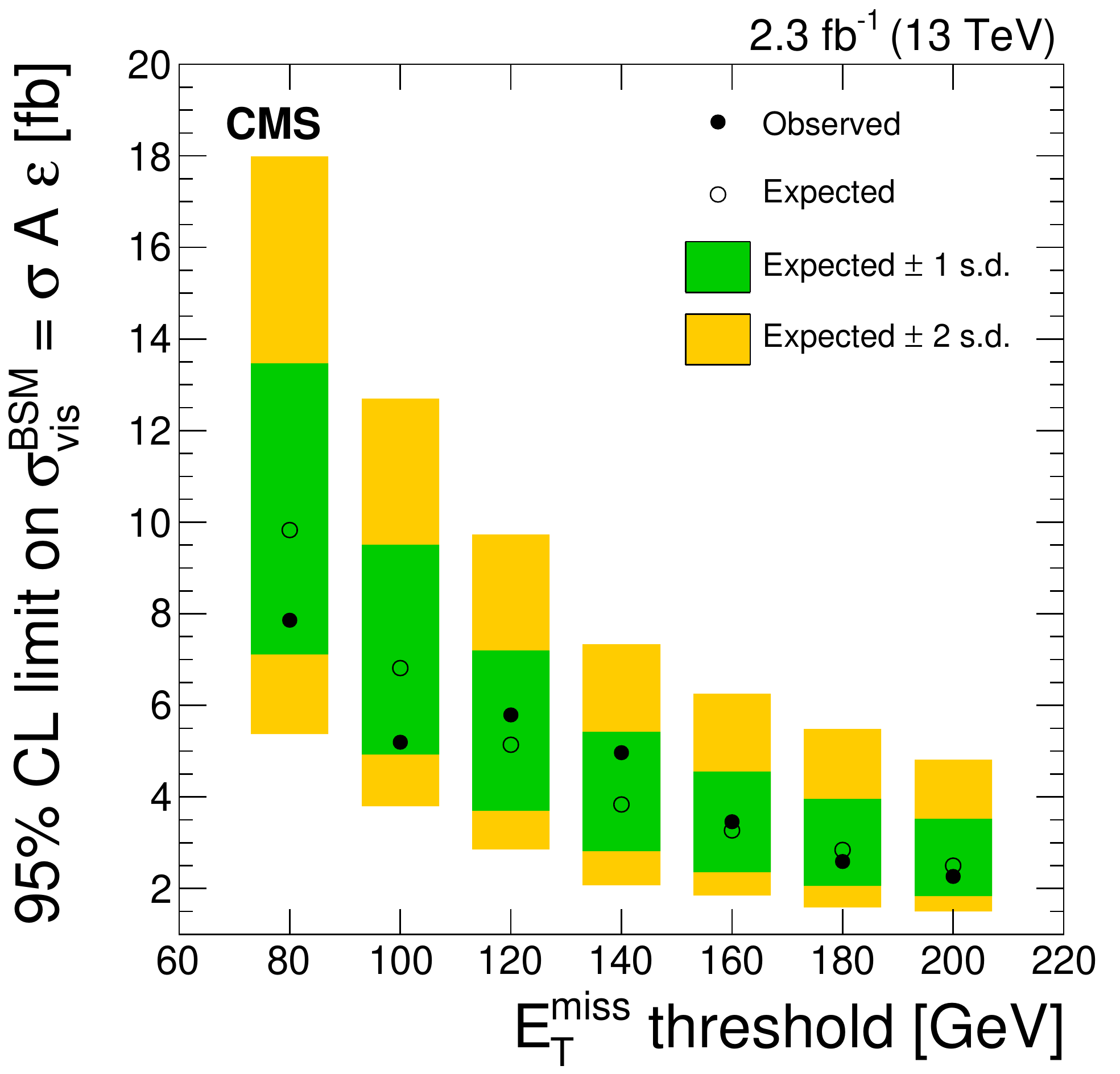}
\caption{The model-independent upper limits at 95\% \CL on the visible cross section ($\sigma\, A\,\epsilon$) for BSM production of events, as a function of \ETm threshold. The values plotted correspond to those given in Table~\ref{tab:model_indep}.}
\label{fig:modelIndepLimits}
\end{figure}

As an alternative to the interpretation of the results in specific models, a simple counting experiment is performed to obtain model-independent
expected and observed 95\% \CL upper limits on the visible cross section $\sigma_\text{vis}^\mathrm{BSM} = \sigma \, A \, \epsilon$
for BSM physics processes, where $A$ is the acceptance and $\epsilon$ is the identification efficiency for a hypothetical signal.
The limits as a function of \ETm thresholds are shown in Fig.~\ref{fig:modelIndepLimits}.
Table~\ref{tab:model_indep} shows the total SM background predictions for the numbers of events passing the selection requirements,
for different \ETm thresholds, compared with the observed numbers of events.
The 95\% \CL expected and observed upper limits for the contribution of events from BSM sources are also shown.
Since the efficiency of reconstructing potential signal events depends on the characteristics of the signal, the model-independent limits are
not corrected for the efficiency. For the models considered in this analysis, typical efficiencies are in the range 50--70\% (simplified DM model), 60--70\% (EFT DM model), and 55--60\% (unparticle model).
The efficiencies are calculated as the ratio of the number of simulated events passing the final selection to the number of simulated events passing the selection criteria at the generator level (acceptance).

\begin{table*}[!htb]
\centering
\topcaption{Total SM background predictions for the numbers of events passing the selection requirements,
        for different \ETm thresholds, compared with the observed numbers of events.
        The listed uncertainties include both statistical and systematic components.
        The 95\% \CL observed and expected upper limits for the contribution of events from BSM sources are also shown.
        In addition, the ${\pm}1$ s.d. and ${\pm}2$ s.d. excursions from expected limits are given.}
\begin{tabular}{lccccccc}
\hline
\ETm threshold [\GeVns{}] & 80 & 100 & 120 & 140 & 160 & 180 & 200\\
\hline
Total SM background	 &  73.9 & 43.0 & 24.0 & 14.1 & 9.5 & 6.8 & 4.9 \\
Total uncertainty  &~~9.2 &~~5.2 &~~2.9 &~~1.3 &  0.9 &  0.7 &  0.5 \\

Data     &   66  &   37  &   26  &   17  &  10  &   6  &   4\\
\hline
Observed upper limit         & 18.1 & 11.9 & 13.3 & 11.4 &~~8.0 &~~6.0 &~~5.2 \\
Expected upper limit $+$2 s.d. & 41.4 & 29.2 & 22.4 & 16.9 & 14.4 & 12.6 & 11.1 \\
Expected upper limit $+$1 s.d. & 31.0 & 21.9 & 16.5 & 12.5 & 10.5 &~~9.1 &~~8.1 \\
Expected upper limit         & 22.6 & 15.7 & 11.8 &~~8.8 &~~7.5 &~~6.5 &~~5.8 \\
Expected upper limit $-$1 s.d. & 16.4 & 11.3 &~~8.5 &~~6.5 &~~5.4 &~~4.7 &~~4.2 \\
Expected upper limit $-$2 s.d. & 12.4 &~~8.7 &~~6.6 &~~4.8 &~~4.3 &~~3.7 &~~3.5 \\
\hline
\end{tabular}
\label{tab:model_indep}
\end{table*}
\clearpage
\section{Summary}

A search for physics beyond the standard model has been performed in events with a \Z
boson and missing transverse momentum, using a data set corresponding to an integrated luminosity of 2.3\fbinv
of $\Pp\Pp$ collisions at a center-of-mass energy of 13\TeV.  The observed data are consistent
with the expected standard model processes.  The results are analyzed to obtain limits in three different scenarios of physics
beyond the standard model.  In a simplified model of DM production
 via a vector or axial vector mediator, 95\% confidence level limits are obtained on the masses of the DM particles and the mediator.
Limits on the DM-nucleon scattering cross section are set at 90\% confidence level in spin-dependent and spin-independent coupling scenarios.
In an effective field theory approach, limits are set on the DM coupling parameters to $U(1)$ and $SU(2)$ gauge fields and on the scale of new physics.
For an unparticle model, 95\% confidence level limits are obtained on the effective cutoff scale as a function of the scaling dimension.
In addition, model-independent limits on the contribution to the visible
$\Z+\MET$ cross section from non-standard-model sources are presented as a function of the minimum requirement on \ETm.
These results are the first in this signal topology to be interpreted in terms of a simplified model. Furthermore, the limits on unparticle production are the first of their kind to be presented at $\sqrt{s}=13\TeV$.

\begin{acknowledgments}
We congratulate our colleagues in the CERN accelerator departments for the excellent performance of the LHC and thank the technical and administrative staffs at CERN and at other CMS institutes for their contributions to the success of the CMS effort. In addition, we gratefully acknowledge the computing centres and personnel of the Worldwide LHC Computing Grid for delivering so effectively the computing infrastructure essential to our analyses. Finally, we acknowledge the enduring support for the construction and operation of the LHC and the CMS detector provided by the following funding agencies: BMWFW and FWF (Austria); FNRS and FWO (Belgium); CNPq, CAPES, FAPERJ, and FAPESP (Brazil); MES (Bulgaria); CERN; CAS, MoST, and NSFC (China); COLCIENCIAS (Colombia); MSES and CSF (Croatia); RPF (Cyprus); SENESCYT (Ecuador); MoER, ERC IUT and ERDF (Estonia); Academy of Finland, MEC, and HIP (Finland); CEA and CNRS/IN2P3 (France); BMBF, DFG, and HGF (Germany); GSRT (Greece); OTKA and NIH (Hungary); DAE and DST (India); IPM (Iran); SFI (Ireland); INFN (Italy); MSIP and NRF (Republic of Korea); LAS (Lithuania); MOE and UM (Malaysia); BUAP, CINVESTAV, CONACYT, LNS, SEP, and UASLP-FAI (Mexico); MBIE (New Zealand); PAEC (Pakistan); MSHE and NSC (Poland); FCT (Portugal); JINR (Dubna); MON, RosAtom, RAS and RFBR (Russia); MESTD (Serbia); SEIDI and CPAN (Spain); Swiss Funding Agencies (Switzerland); MST (Taipei); ThEPCenter, IPST, STAR and NSTDA (Thailand); TUBITAK and TAEK (Turkey); NASU and SFFR (Ukraine); STFC (United Kingdom); DOE and NSF (USA).

Individuals have received support from the Marie-Curie programme and the European Research Council and EPLANET (European Union); the Leventis Foundation; the A. P. Sloan Foundation; the Alexander von Humboldt Foundation; the Belgian Federal Science Policy Office; the Fonds pour la Formation \`a la Recherche dans l'Industrie et dans l'Agriculture (FRIA-Belgium); the Agentschap voor Innovatie door Wetenschap en Technologie (IWT-Belgium); the Ministry of Education, Youth and Sports (MEYS) of the Czech Republic; the Council of Science and Industrial Research, India; the HOMING PLUS programme of the Foundation for Polish Science, cofinanced from European Union, Regional Development Fund, the Mobility Plus programme of the Ministry of Science and Higher Education, the National Science Center (Poland), contracts Harmonia 2014/14/M/ST2/00428, Opus 2013/11/B/ST2/04202, 2014/13/B/ST2/02543 and 2014/15/B/ST2/03998, Sonata-bis 2012/07/E/ST2/01406; the Thalis and Aristeia programmes cofinanced by EU-ESF and the Greek NSRF; the National Priorities Research Program by Qatar National Research Fund; the Programa Clar\'in-COFUND del Principado de Asturias; the Rachadapisek Sompot Fund for Postdoctoral Fellowship, Chulalongkorn University and the Chulalongkorn Academic into Its 2nd Century Project Advancement Project (Thailand); and the Welch Foundation, contract C-1845.
\end{acknowledgments}

\bibliography{auto_generated}

\cleardoublepage \appendix\section{The CMS Collaboration \label{app:collab}}\begin{sloppypar}\hyphenpenalty=5000\widowpenalty=500\clubpenalty=5000\textbf{Yerevan Physics Institute,  Yerevan,  Armenia}\\*[0pt]
A.M.~Sirunyan, A.~Tumasyan
\vskip\cmsinstskip
\textbf{Institut f\"{u}r Hochenergiephysik,  Wien,  Austria}\\*[0pt]
W.~Adam, E.~Asilar, T.~Bergauer, J.~Brandstetter, E.~Brondolin, M.~Dragicevic, J.~Er\"{o}, M.~Flechl, M.~Friedl, R.~Fr\"{u}hwirth\cmsAuthorMark{1}, V.M.~Ghete, C.~Hartl, N.~H\"{o}rmann, J.~Hrubec, M.~Jeitler\cmsAuthorMark{1}, A.~K\"{o}nig, I.~Kr\"{a}tschmer, D.~Liko, T.~Matsushita, I.~Mikulec, D.~Rabady, N.~Rad, B.~Rahbaran, H.~Rohringer, J.~Schieck\cmsAuthorMark{1}, J.~Strauss, W.~Waltenberger, C.-E.~Wulz\cmsAuthorMark{1}
% ...
\vskip\cmsinstskip
% ...
1:~~Also at Vienna University of Technology, Vienna, Austria\\
% ...

\end{sloppypar}
\end{document}